\documentclass[twocolumn]{aastex7}

\usepackage{booktabs} 
\usepackage{array} 
\usepackage{xcolor}

\begin{document}

\title{The dawn is quiet II: Gaia XP constraints on the Milky Way's proto-Galaxy\\ from very metal-poor MDF tails}

\shorttitle{The dawn is quiet II}

\author[0000-0001-7083-2417]{Boquan Chen} 
\affiliation{Department of Astronomy, The Ohio State University, Columbus, OH 43210, USA}
\affiliation{Center for Cosmology and AstroParticle Physics (CCAPP), 191 West Woodruff Avenue, Columbus, OH 43210, USA}
\email{ebchen.astro@gmail.com}

\author[0000-0002-8151-8498]{Matthew Orkney} 
\affiliation{Institut de Ciencies del Cosmos (ICCUB), Universitat de Barcelona (IEEC-UB), Martí i Franquès 1, E08028 Barcelona, Spain}
\affiliation{Institut d'Estudis Espacials de Catalunya (IEEC), E-08034 Barcelona, Spain}
\email{morkney@icc.ub.edu}

\author[0000-0001-5082-9536]{Yuan-Sen Ting} 
\affiliation{Department of Astronomy, The Ohio State University, Columbus, OH 43210, USA}
\affiliation{Center for Cosmology and AstroParticle Physics (CCAPP), 191 West Woodruff Avenue, Columbus, OH 43210, USA}
\email{ting.74@osu.edu}

\author[0000-0001-7294-9766]{Michael Hayden} 
\affiliation{Department of Physics and Astronomy, The University of Oklahoma, 440 W. Brooks St., Norman, OK 73019, USA}
\email{mrhayden@ou.edu}

\begin{abstract}
The earliest phase of the Milky Way's evolution involved a transition from a dispersion-supported proto-galaxy to a rotationally supported disk. A key chemical signature of this transition is the moderate rise in [$\alpha$/Fe] near $\mathrm{[Fe/H]}\approx-1.3$, which we previously interpreted as evidence for $\alpha$-enhanced gas accretion fueling early disk formation. However, this trend alone does not uniquely constrain the trade-off between initial gas mass, inflow rate, and star formation efficiency (SFE), leaving the physical condition of the proto-Milky Way uncertain. To break this degeneracy, we analyze the metal-poor tail ($-3<\mathrm{[Fe/H]}<-2$) of the Milky Way's metallicity distribution function (MDF) using Gaia DR3 BP/RP (XP) metallicities from ten catalogs. After applying recommended quality cuts, all catalogs exhibit a single-slope exponential tail with slopes $k\simeq0.6$--2.0. Comparison with one-zone galactic chemical-evolution (GCE) models that replicated the [$\alpha$/Fe]-rise from Paper I shows that shallow tails ($k\simeq0.6$) require a massive initial cold gas reservoir ($\gtrsim10^9\, \mathrm{M_\odot}$), while steeper tails ($k\gtrsim1$) arise from small reservoirs that built up over time with weak inflow. MDFs with $k \simeq 1.0$ are best reproduced under our GCE framework, which favor a proto-Galaxy with a moderate gas reservoir ($10^{8}$--$10^{9}\, \mathrm{M_\odot}$) sustained through weak continuous inflow ($\sim 2 \ \mathrm{M_\odot \ yr^{-1}}$) and SFE comparable to today's value (a few $\times 10^{-10}\,\mathrm{yr^{-1}}$) during the first Gyr. This scenario is reinforced by MDFs of 30 Milky Way analogs in the Auriga simulations, which exhibit similar slopes ($k\approx1.25$). The metal-poor MDF tail thus provides a quantitative constraint on the Milky Way's early gas accretion and star formation history.

\end{abstract}

\keywords{\uat{Galaxies}{573}}

\section{Introduction} 

The earliest phase of the Milky Way's formation was dominated by a rapid and turbulent phase in the form of a compact proto-galaxy, preceding the emergence of the rotationally supported disk. Elemental abundances from APOGEE \citep{2017AJ....154...94M} and kinematics from Gaia \citep{2016A&A...595A...1G} indicate an in-situ, kinematically hot ``Aurora'' phase at low metallicity, followed by a fast ``spin-up'' of the Galaxy as its disk started to form between $-1.3 < [\mathrm{Fe/H}] < -0.9$ over $\sim$1--2 Gyr, with a simultaneous drop in abundance scatter \citep{2022MNRAS.514..689B}. A central concentration of 18,000 giant stars with [Fe/H] $< -1.5$ have been found within $\sim 5$ kpc of the Galactic center and proposed to be the proto-Galactic nucleus. These stars suggest that the metal-poor ``heart'' contains $>10^{8} \ \mathrm{M_\odot}$ and formed $>12.5$ Gyr ago \citep{2022ApJ...941...45R}. Multiple components belonging to the proto-Galaxy have also been identified and characterized with a mix of in situ vs accreted origins \citep{2024ApJ...964..104M, 2025MNRAS.537.3730H}. 

The disk of the Milky Way commenced its assembly around 13 Gyr ago, with its star formation rate (SFR) peaking around 11 Gyr ago, coinciding with the Gaia-Sausage-Enceladus (GSE) \citep{2022Natur.603..599X, 2021MNRAS.508.5903V, 2022MNRAS.514..689B, 2025NatAs...9..101X}. This used to be considered unusually early, but a cold rotating disk has been confirmed by an extra-galactic observation at $z \sim 4.3$ when the universe was only 1.5 Gyr old \citep{2020Natur.581..269N}. The quiescent evolution of the disk took over around 8 Gyr ago and continued to the present as evidenced by a clear division in the stellar age-metallicity distribution \citep{2013A&A...560A.109H, 2015A&A...578A..87S, 2019MNRAS.489.1742F, 2022MNRAS.510.4669S}. These three phases of evolution are confirmed by a sample of ten million giants from Gaia \citep{2024ApJ...972..112C}. However, there is evidence that the kinematically defined thin and thick disk could have also co-evolved from 13 Gyr ago \citep{2024A&A...688A.167N, 2025arXiv250400135B}. 

As the Galactic disk began to form, a moderate rise in [$\alpha$/Fe] appeared near $\mathrm{[Fe/H]} \approx -1.3$, even among stars on low-eccentricity orbits \citep{2022arXiv220402989C, 2022MNRAS.514..689B, 2022ApJ...941...45R}. This feature marks the transition from the turbulent, kinematically hot Aurora phase to the onset of ordered, rotationally supported star formation and has been interpreted as the chemical signature of the disk's birth. In Paper I \citep{2024PASA...41...63C}, we showed that the observed [$\alpha$/Fe] upturn is best explained by the sudden onset of $\alpha$-enhanced gas inflow-pre-enriched by core-collapse supernovae in the halo-which supplied both angular momentum and fuel for renewed star formation after a preceding inflow-poor phase. The resulting enhancement thus reflects the coupled chemical and dynamical effects of the accreted material. However, the  [$\alpha$/Fe]-rise alone cannot uniquely determine the proto-Galaxy's initial gas mass, star-formation efficiency, or inflow timescale, as multiple parameter combinations yield similar patterns. To break this degeneracy, we now turn to the slope of the metal-poor tail of the metallicity distribution function (MDF), which traces the balance between gas supply and metal production during the Galaxy's earliest evolution.


Very metal-poor (VMP; $\mathrm{[Fe/H]} < -2.0$) and extremely metal-poor (EMP; $\mathrm{[Fe/H]} < -3.0$) stars provide the most abundant fossil record of the proto-Galaxy, retaining the chemical and kinematic signatures of the first star-formation episodes. Even small, well-characterized samples offer strong constraints on early enrichment and gas accretion-for instance, asteroseismic ages of $\sim$ 12 Gyr for stars with $\mathrm{[Fe/H]} \approx -2.7$ set lower bounds on the onset of in-situ disk formation \citep{2024ApJ...975...19H}. Interpreting such ancient populations requires GCE modeling, which links observables such as the MDF, [$\alpha$/Fe]-[Fe/H] trends, and age-abundance relations to the underlying inflow, outflow, and star formation histories (SFH). Historically, VMP and EMP stars have constrained the critical metallicity for low-mass star formation and the yields of the first supernovae \citep{1976ApJ...209..418H, 2003A&A...404..211P, 2007MNRAS.381..647S, 2013A&A...554A.135B}, but robust statistical inference has been limited by the rarity of such stars and strong selection biases. {Gaia DR3 BP/RP spectrophotometry (XP spectra) \citep{2023A&A...674A...1G, 2023A&A...674A...2D} now makes it possible to measure the metal-poor tail of the Galactic MDF with homogeneous, all-sky stellar-parameter catalogs far larger than any previous spectroscopic sample.}


{Modern cosmological simulations provide a physically grounded framework for interpreting observational constraints on early disk assembly. The FIRE Project \citep{2014MNRAS.445..581H} is a suite of high-resolution zoom simulations that resolves stellar feedback processes and their impact on gas regulation. The Auriga simulations \citep{10.1093/mnras/stx071} follow Milky Way–mass galaxies in a $\Lambda$CDM context with detailed chemo-dynamical evolution and realistic disk formation. IllustrisTNG \citep{2019MNRAS.490.3196P} provides large-volume statistics linking gas accretion, enrichment, and stellar mass growth across galaxy populations.} Across these frameworks, Milky Way–mass progenitors at $z\gtrsim2$ are predicted to experience bursty star formation fueled by cold, dense inflows \citep{2006MNRAS.368....2D, 2015MNRAS.454.2691M, 2021ApJ...911...88S}, with feedback-driven outflows regulating early enrichment and driving strong time variability in the star-formation history \citep{2014MNRAS.445..581H}. 

Although early star formation is centrally concentrated, repeated burst–feedback cycles can dynamically heat stellar orbits and disperse the oldest populations into spheroidal, halo-like configurations \citep{2016ApJ...820..131E,2018MNRAS.480..652E}. The resulting proto-galaxies are typically built by one or two dominant progenitors plus a small number of lower-mass companions \citep{2024MNRAS.527.9810H}. In Auriga, the early stellar component remains compact and metal-poor, with mild net rotation and $\alpha$-enhanced chemistry broadly resembling the oldest Galactic disk populations, while continued cosmological inflow sustains star formation and generates extended metal-poor populations \citep{10.1093/mnras/stx071}. The transition to an ordered, rotationally supported stellar disk occurs as inflow and feedback-driven turbulence become less disruptive due to the change in the gravitational potential. Once gas accretion stabilizes, a coherent disk can emerge within the first $\sim$1--2 Gyr of cosmic time and, in some cases, assemble rapidly (within $\sim$0.2 Gyr) despite residual burstiness, attaining the rotation and $\alpha$-enhanced chemical patterns characteristic of the Milky Way’s most metal-poor disk stars \citep{2021MNRAS.505..889Y,2023MNRAS.519.2598G, 2023MNRAS.525.2241H, 2025ApJ...990....7S}.


In this paper, we use Gaia DR3 XP metallicities to quantify the metal-poor tail of the Milky Way's MDF over $-3<\mathrm{[Fe/H]}< -2.0$ and interpret the measured slopes within a one-zone GCE framework. Building on Paper I, which linked the observed rise in [$\alpha$/Fe] at $\mathrm{[Fe/H]} \approx -1.3$ to the onset of $\alpha$-enhanced inflow and disk spin-up, we now constrain the proto-Galaxy's parameters using a complementary diagnostic: the logarithmic slope of the metal-poor MDF tail. This approach leverages Gaia XP's homogeneous, Galaxy-wide sampling of VMP stars to test whether the Milky Way's earliest phase resembled the burst-regulated, gas-rich progenitors predicted by cosmological simulations. Together, these analyses aim to provide a quantitative, physically motivated description of the conditions under which the Galactic disk first emerged.

{This paper is organized as follows. Section~\ref{sec:data} summarizes the Gaia DR3 XP metallicity catalogs used in this work and the adopted quality cuts, and presents basic CMD and spatial/kinematic diagnostics to characterize their selection footprints and potential contamination. Section~\ref{sec:gce} describes the GCE model grid (introduced in Paper~I) and the procedure for predicting and measuring metal-poor MDF tail slopes. In Section~\ref{sec:results} we measure the logarithmic MDF slopes over $-3<\mathrm{[Fe/H]}<-2.0$ across catalogs, test their robustness, and interpret the inferred $k$ values in terms of the proto-Galaxy’s parameters. We then compare the observed slopes to Milky Way analogs in the Auriga simulations. Section~\ref{sec:discussion} discusses systematic uncertainties and the connection to prior halo MDF and GCE studies. Section~\ref{sec:conclusion} summarizes our conclusions.}

\section{Gaia XP metallicity catalogs}
\label{sec:data}

\subsection{Before Gaia}

Before Gaia DR3, assembling large and homogeneous samples of very metal-poor stars relied on a sequence of survey strategies that progressively expanded both sky coverage and statistical accuracy. {Early objective-prism work demonstrated that the Ca\,II H\&K resonance lines provide a robust metallicity diagnostic at low spectral resolution; in particular, \citet{1984ApJ...285..622B} identified a star with $[\mathrm{Fe/H}] = -4.5$, establishing the feasibility of systematically probing the metal-poor tail. Subsequent discoveries pushed the metallicity frontier to the ultra-metal-poor regime ($\mathrm{[Fe/H]} < -4$) and revealed surprising findings on their elemental abundances \citep{2002Natur.419..904C, 2005Natur.434..871F, 2011Natur.477...67C}.} 

This approach was scaled to large-area surveys such as the HK survey \citep{1985AJ.....90.2089B, 1992AJ....103.1987B} and the Hamburg/ESO (HES) survey \citep{2000A&A...358...77W, 2008A&A...484..721C}, transforming isolated discoveries into statistically meaningful halo samples. Medium-resolution spectroscopic campaigns, including SEGUE \citep{2009AJ....137.4377Y} and RAVE \citep{2007MNRAS.379..755S, 2017A&A...603A..19M}, further expanded the census of metal-poor stars across wide sky areas. Complementary photometric programs such as SkyMapper \citep{2007PASA...24....1K} and Pristine \citep{2017MNRAS.471.2587S} employed metallicity-sensitive filters for efficient candidate pre-selection, while LAMOST \citep{2012RAA....12..723Z} provided extensive low-resolution spectroscopy from which metal-poor targets could be drawn. Compiled databases such as SAGA \citep{2008PASJ...60.1159S} and JINAbase \citep{2018ApJS..238...36A} aggregated high-resolution abundance measurements, though with heterogeneous selection functions. Together, these efforts established the principal methodologies for identifying metal-poor stars—low-resolution spectral indicators of line blanketing, photometric metallicity proxies, and broad spectroscopic classification—and they form the historical foundation against which the homogeneous Gaia XP metallicity measurements can be interpreted.

\subsection{Training Methods for Gaia Catalogs}

{The Gaia XP spectra, released in DR3, provide low-resolution ($R \sim 20$--$100$) spectrophotometry from the BP and RP instruments covering 330--1050 nm for sources down to $G \approx 17.65$ \citep{2023A&A...674A...2D}.} Although the resolving power is significantly lower than that of most spectroscopic surveys, the all-sky coverage, homogeneous calibration, and continuous spectral sampling encode metallicity information for tens of millions of stars, including very metal-poor populations. XP shares methodological features with earlier objective-prism and photometric surveys—both rely on broad-band spectral energy distributions and weak metal-line signatures—but with the advantage of contiguous spectrophotometric data and a uniform instrument response.

We provide a brief summary of the various techniques used to derive metallicities from Gaia XP spectra here. The Gaia DR3 pipeline serves as the baseline of analyzing the XP spectra (see GaiaXPy\footnote{\url{https://www.cosmos.esa.int/web/gaia/gaiaxpy}} for more details). GSP-Phot fits XP spectra plus parallax and $G$ to synthetic libraries in a Bayesian forward model, producing $T_{\mathrm{eff}}$, $\log\, g$, and a calibrated metallicity for $\sim 471\,\mathrm{M}$ sources (useful mostly qualitatively for $[\mathrm{M}/\mathrm{H}]$ per their own validation) \citep{2023A&A...674A..27A}. The signal-to-noise (S/N) ratio of the XP coefficients is a key factor in the reliability of these estimates. Building on that, \cite{2023ApJS..267....8A} trained gradient-boosted trees on APOGEE DR17 labels augmented with a curated very metal-poor set, and fed the model XP, Gaia, broadband, narrowband (including synthetic Pristine CaHK) and CatWISE photometry \citep{2021ApJS..253....8M} as features; they predict $T_{\mathrm{eff}}$, $\log\, g$, $[\mathrm{M}/\mathrm{H}]$ for $\sim 175\,\mathrm{M}$ stars and also provide a carefully vetted bright RGB subset ($G \lesssim 16$) where the metallicity is most reliable.

SHBoost \citep{2024A&A...691A..98K} targets essentially ``all stars with XP and parallaxes'': it learns $T_{\mathrm{eff}}$, $\log\,g$, $[\mathrm{M}/\mathrm{H}]$, $A_V$, and mass with XGBoost trained on StarHorse labels compiled from many spectroscopic surveys (plus curated hot/VMP/WD sub-samples) \citep{2018MNRAS.476.2556Q} using XP coefficients together with Gaia+2MASS\citep{2006AJ....131.1163S}+WISE\citep{2010AJ....140.1868W}; the team also applies a posteriori calibration for metallicity systematics. \citet{2023MNRAS.524.1855Z} go the other way—forward-modelling XP and broadband photometry with a data-driven model trained on LAMOST; they provide $T_{\mathrm{eff}}$, $\log\, g$, $[\mathrm{Fe}/\mathrm{H}]$ (and revised distances/extinctions) for $\sim 220\,\mathrm{M}$ stars, i.e., an all-sky, all-types sample anchored to LAMOST labels.

Several recent XP-based analyses apply more targeted strategies, either by improving the XP flux calibration before modeling or by restricting the training domain to regimes where the abundance information is strongest. \citet{2025A&A...695A..75Y} first infer per-star flux corrections with an APOGEE-trained neural network to suppress small-scale XP wiggles and then fit the corrected spectra with \texttt{FERRE} against synthetic grids, producing a $68\,\mathrm{M}$-star catalog of $T_{\mathrm{eff}}$, $\log\, g$, and $[\mathrm{M}/\mathrm{H}]$ with enhanced control of systematics across the sky. AspGap \citep{2024ApJS..272....2L} focuses on red-giant branch stars and employs a neural network trained on APOGEE that includes an auxiliary component designed to reconstruct the corresponding high-resolution APOGEE spectrum. This additional reconstruction objective strengthens the link between XP features and abundance-sensitive spectral structure, yielding $T_{\mathrm{eff}}$, $\log\, g$, $[\mathrm{M}/\mathrm{H}]$, and precise $[\alpha/\mathrm{M}]$ for $\sim 37\,\mathrm{M}$ RGB stars.

\citet{2025ApJ...980...90H} adopt a complementary strategy optimized for low-extinction sightlines. They apply explicit cuts in Galactic latitude and $E(B-V)$, impose a $G < 18.85$ limit, and train boosted trees on APOGEE DR17 augmented with a curated very metal-poor sample, publishing $[\mathrm{M}/\mathrm{H}]$ and $[\alpha/\mathrm{M}]$ for $\sim 48\,\mathrm{M}$ stars in these dust-minimal windows. \citet{2025ApJS..279....7Y} emphasize giants as well, dereddening and correcting XP spectra before using an uncertainty-aware, cost-sensitive neural network (UA-CSNet) trained on high-quality spectroscopic labels (PASTEL) to predict $[\mathrm{Fe}/\mathrm{H}]$ for $\sim 20\,\mathrm{M}$ giants with particular attention to the metal-poor tail.

Two efforts push specific elemental abundances beyond metallicity. \citet{2025MNRAS.537.1984A} train a neural network on spectroscopic reference labels to infer both $[\mathrm{Fe}/\mathrm{H}]$ and $[\mathrm{C}/\mathrm{Fe}]$ from XP (plus auxiliary Gaia information) for $>10\,\mathrm{M}$ stars, then identify $\sim 2{,}000$ bright carbon-enhanced metal-poor (CEMP) candidates; their tables link back to vetted giant selections to ensure reliability at the metal-poor end. \citet{2025MNRAS.536.2507K} focus on abundance patterns associated with ``second-generation'' globular cluster (GC) populations. They employ a multilayer perceptron (MLP) neural network trained on APOGEE cross-matches to predict $T_{\mathrm{eff}}$, $\log\, g$, $[\mathrm{Fe}/\mathrm{H}]$, $[\mathrm{N}/\mathrm{O}]$, and $[\mathrm{Al}/\mathrm{Fe}]$ directly from normalized XP coefficients. Applied to the Andrae-vetted RGB sample, these predictions allow them to identify high-$[\mathrm{N}/\mathrm{O}]$ field-star candidates that may trace a GC-origin channel. Finally, one catalog uses XP to synthesize a metallicity-sensitive narrow band: the Pristine DR1-XP release computes synthetic CaHK magnitudes for $\sim 219\,\mathrm{M}$ stars and, combined with broadband Gaia (and deeper Pristine where available), publishes photometric metallicities for $>30\,\mathrm{M}$ high-S/N FGK stars, with strong performance below $[\mathrm{Fe}/\mathrm{H}] \approx -1$ and extending to $\lesssim -3.5$ \citep{2024A&A...692A.115M}.

Most XP metallicity catalogs are supervised and anchored to big spectroscopic surveys (APOGEE, LAMOST, StarHorse), but they differ in who they target (all-sky vs.\ giants vs.\ low-extinction), how they treat the XP fluxes (raw vs.\ corrected vs.\ forward-modelled), and which labels they predict (some give $[\mathrm{M}/\mathrm{H}]$, some $[\mathrm{Fe}/\mathrm{H}]$, many now add $[\alpha/\mathrm{M}]$ or $[\mathrm{C}/\mathrm{Fe}]$). These choices set the reliability envelope depending on the specific user case. We compile in Table~\ref{tab:catalog_summary} a summary of the major public catalogs derived from Gaia XP spectra mentioned above. The table lists the size of each release, the spectroscopic or synthetic training sets, the modeling framework (e.g., forward models, tree-based regressors, or neural networks), and the authors' recommended quality cuts that constrain each catalog to its validated regime which we adopt for our analysis. This overview is intended to clarify the heterogeneity among XP analyses and to guide the selection of appropriate subsamples for metal-poor studies.

{Because the Gaia XP metallicity catalogs do not share a common selection function, we use the published metallicities primarily for differential analyses. We do not attempt to construct a volume-complete sample or combine all catalogs into a single homogenized dataset, but instead treat each catalog within its published quality-controlled regime. Several potential biases are relevant. First, metal-poor giants are intrinsically brighter than dwarfs, so magnitude-limited samples can over-represent giants and probe different Galactic volumes and component mixtures depending on the CMD coverage of each catalog. Second, because the Gaia XP samples span both disk and halo sightlines, contamination by metal-rich disk stars with underestimated metallicities may affect the metal-richer boundary of the inferred metal-poor tail. Since these effects depend on the training set, inference domain, and recommended quality cuts of each public release, they must be evaluated empirically when interpreting the MDFs (see Section 4.2).}

\begin{deluxetable*}{l l l l l}
\tabletypesize{\footnotesize}
\tablecaption{Summary of Stellar Parameter Catalogs from Gaia XP Spectra}
\label{tab:catalog_summary}
\tablehead{
\colhead{Reference} &
\colhead{Number of Stars\tablenotemark{a}} &
\colhead{Training Data} &
\colhead{Method} &
\colhead{Adopted Recommended Quality Cuts}
}
\startdata
Andrae et al. (2023) &
222{,}572 &
\parbox[t]{4.5cm}{APOGEE DR17 + VMP stars (Li et al. 2022)} &
\parbox[t]{4.5cm}{{XGBoost}: Gradient-boosted trees with weighted training.} &
\parbox[t]{6.5cm}{Use the provided bright RGB sample ($G<16$) vetted for high precision and purity. } \\
Zhang et al. (2023) &
3{,}326{,}913 &
\parbox[t]{4.5cm}{LAMOST DR8 + ``Hot Payne''} &
\parbox[t]{4.5cm}{{Empirical Forward Model}: NN predicts spectra from labels, which are then inferred.} &
\parbox[t]{6.5cm}{Use the \texttt{quality\_flags} column ($<8$), incorporating cuts on $\chi^2$, prior probability, and XP–parallax consistency. We further restrict to stars within the training-set convex hull: $3500 < T_\mathrm{eff} < 8500$, $-0.5 < \log\,g < 5$.} \\
Li et al. (2023) &
155{,}145 &
\parbox[t]{4.5cm}{APOGEE DR17} &
\parbox[t]{4.5cm}{{Neural Network (AspGap)}: Includes a ``hallucinator'' component.} &
\parbox[t]{6.5cm}{Follow the recommended red-giant-only sample ($\sim 12$ million stars). We also adopt $\sigma_{\mathrm{[M/H]}}<0.15$ dex.} \\
Martin et al. (2024) &
1{,}785{,}328 &
\parbox[t]{4.5cm}{Spectroscopic sample (SDSS, APOGEE) + Synthetic spectra (MARCS)} &
\parbox[t]{4.5cm}{{Photometric Grid Model}: Calibrated color–metallicity grid using synthetic CaHK magnitudes.} &
\parbox[t]{6.5cm}{Use flag \texttt{mcfrac > 0.8} to ensure Ca H\&K magnitude is computed from valid Gaia XP coverage, \texttt{Pvar < 0.3}, \texttt{RUWE < 1.4}.} \\
Fallows \& Sanders (2024) &
233{,}605 &
\parbox[t]{4.5cm}{APOGEE DR17} &
\parbox[t]{4.5cm}{{Uncertain Neural Network}: NN with explicit modeling of multiple uncertainty components.} &
\parbox[t]{6.5cm}{No separate quality flag; only sources with $4000< T_\mathrm{eff}<7000$ K and $0<\log g<5$ appear in the released catalog.} \\
Hattori (2024) &
1{,}447{,}280 &
\parbox[t]{4.5cm}{APOGEE DR17 + VMP stars} &
\parbox[t]{4.5cm}{{Quantile Regression Forests (QRF)}: Tree-based ensemble predicting full probability distributions.} &
\parbox[t]{6.5cm}{Restrict to low-extinction sources with $E(B-V)<0.1$. Use the \texttt{bool\_flag\_cmd\_good} flag to select stars within the trained CMD domain.} \\
Khalatyan et al. (2024) &
279{,}126 &
\parbox[t]{4.5cm}{StarHorse (Q23; compilation of many spectroscopic surveys)} &
\parbox[t]{4.5cm}{{XGBoost (SHBoost)}: Uses \texttt{xgboost-distribution} for uncertainty estimation.} &
\parbox[t]{6.5cm}{Use the output quality flag (\texttt{xgb\_met\_outputflag = 0}) and magnitude cut ($G\leq16.5$).} \\
Kane et al. (2024) &
80{,}696 &
\parbox[t]{4.5cm}{APOGEE DR17} &
\parbox[t]{4.5cm}{{Neural Network (MLP)}: Heteroscedastic regression to classify GC-origin stars.} &
\parbox[t]{6.5cm}{No specific quality cuts for the released catalog. Published sources satisfy S/N $>30$, $4000<T_{\mathrm{eff}}<6500$ K, and $>99\%$ GC membership.} \\
Ardern-Arentsen et al. (2025) &
92{,}352 &
\parbox[t]{4.5cm}{LAMOST spectra re-analysed with \texttt{FERRE} + high-resolution VMP/CEMP literature; XP coefficients as inputs.} &
\parbox[t]{4.5cm}{{Neural Network (heteroscedastic)}: ANN on normalized XP coefficients; outputs $T_{\mathrm{eff}}$, $\log g$, [Fe/H], [C/Fe] with per-star variances; ensemble-averaged predictions.} &
\parbox[t]{6.5cm}{Recommended ``clean bright'' sample: $G<16$, $E(B{-}V)<0.3$, $\mathrm{Pvar}<0.5$, $(\mathrm{BP{-}RP})_0<1.35$.} \\
Ye et al. (2025) &
712{,}821 &
\parbox[t]{4.5cm}{APOGEE DR17 (for flux correction model)} &
\parbox[t]{4.5cm}{{Hybrid}: NN for spectral correction, then $\chi^2$ fitting to synthetic spectra.} &
\parbox[t]{6.5cm}{Use the \texttt{dflux\_per} flag ($<0.08$). Recommended temperature range is $4000~\mathrm{K} \le T_{\mathrm{eff}} \le 7000~\mathrm{K}$. We restrict to $\log\chi^2 < 4$ quality cuts for precise metallicities.} \\
Yang et al. (2025) &
1{,}888{,}361 &
\parbox[t]{4.5cm}{PASTEL catalog} &
\parbox[t]{4.5cm}{{Uncertainty-aware Cost-Sensitive Neural Network (UA-CSNet)}.} &
\parbox[t]{6.5cm}{The reliable sample uses $E(B-V)<0.5$. The model is for giants with $0.7<(\mathrm{BP{-}RP})_0<1.9$. Additional CMD cut with $M_G < 3.5 + 2.5 (\mathrm{BP{-}RP})$. } \\
\enddata

\tablenotetext{a}{With [Fe/H] $<-1.3$ after the adopted quality cuts.}

\end{deluxetable*}

\section{Galactic chemical evolution model}
\label{sec:gce}
We use the same 735 GCE models that successfully replicated the moderate rise in [Mg/Fe] as seen in the H3 survey \citep{2019ApJ...883..107C} from our Paper I. Our model framework is designed to investigate the early [$\alpha$/Fe] behavior in the proto-Milky Way, simulating the first two Gyr of its evolution within a bin with 1 kpc radius. The model divides this period into two one-Gyr phases, with time steps of 30 Myr, reflecting the longest core-collapse supernovae (CCSNe) progenitor lifespans. Stellar populations, ranging from 0.5 to 40 M$_\odot$ and following an initial mass function (IMF) of $\xi(m) \propto m^{-2.3}$, are tracked alongside cold and warm interstellar medium (ISM) components. Nucleosynthetic yields from \citet{2013ARA&A..51..457N} are interpolated across metallicities to determine the chemical enrichment from stars. Stars between 0.5 and 0.9 M$_\odot$ are formed and store cold ISM, but they don't contribute to nucleosynthesis due to the lack of nucleosynthesis yields.

The model incorporates key astrophysical processes. Star formation is governed by the Kennicutt-Schmidt law, converting cold ISM into new stellar populations. Stellar evolution dictates the return of enriched gas: CCSNe from stars more massive than 9 M$_\odot$ explode within one time step (30 Myr), while intermediate-mass stars (3--8 M$_\odot$) evolve into white dwarfs (WDs) that can later produce Type Ia supernovae (SNe Ia). The SNe Ia rate depends on the WD mass reservoir, a constant fraction ($f_{\text{SNIa}}$), and an exponential delay timescale. Enriched material ejected from stars is distributed primarily to the warm ISM (79\%) and a smaller fraction to the cold ISM (1\%), with rest lost to the circum-galactic medium (CGM). The warm ISM cools to replenish the cold ISM, while feedback from star formation and supernovae heats a portion of the cold ISM into the warm ISM, regulated by mass-loading factors, $\eta_\mathrm{SF} = 2.0$ and $\eta_\mathrm{SN} = 5.0$. The outflow mass-loading factors are fixed and directly proportional to the mass involved in star formation and stellar evolution activities. The model also allows for pristine gas inflow into the warm ISM.

To understand the conditions leading to an early [$\alpha$/Fe] rise, we explore a range of five fundamental free parameters: the initial cold ISM mass ($m_{0, \text{cold}}$), the SNe Ia fraction ($f_{\text{SNIa}}$), the warm ISM cooling timescale ($t_{\text{cool}}$), the star formation efficiency ($\epsilon_{\text{SF}}$), and the gas inflow rate ($\dot{m}_{\text{inflow}}$). These parameters are varied across astrophysically motivated ranges in numerous model runs. Notably, to achieve an [$\alpha$/Fe] rise after an initial decline (due to SNe Ia), the star formation efficiency and inflow rate, along with inflow gas composition, are allowed to change in the second Gyr of the simulation, facilitating increased $\alpha$-element production through enhanced CCSNe activity or accretion of $\alpha$-enriched gas.

\section{Results}
\label{sec:results}

\begin{figure*}[pht!]
\plotone{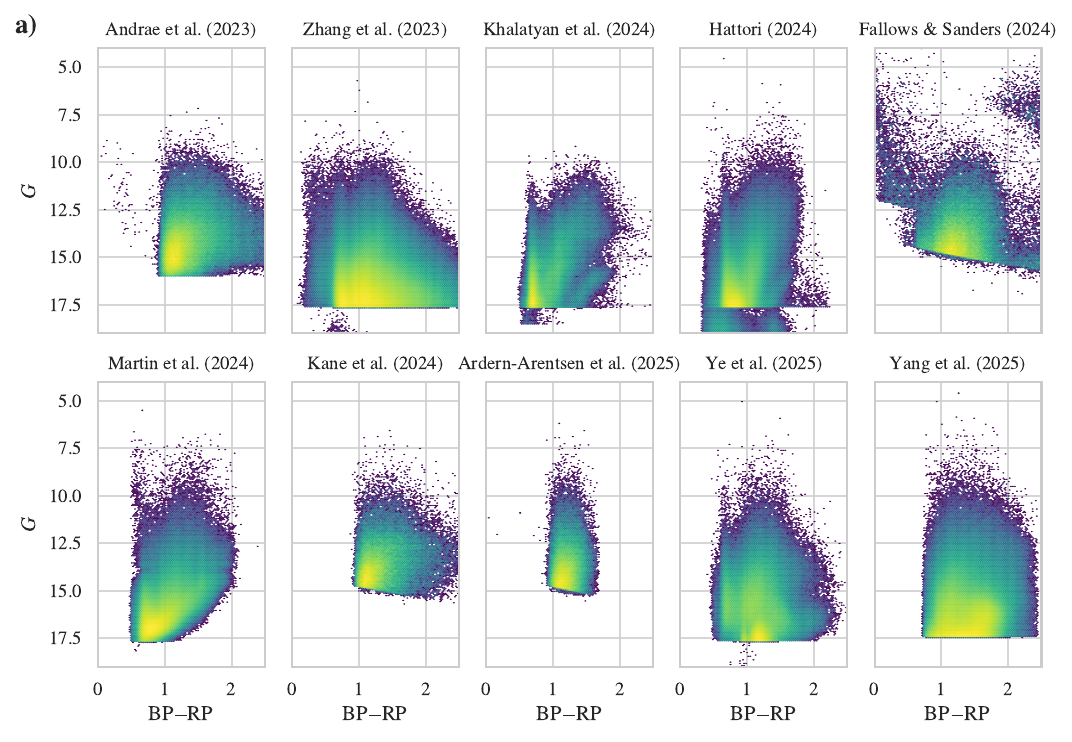}
\plotone{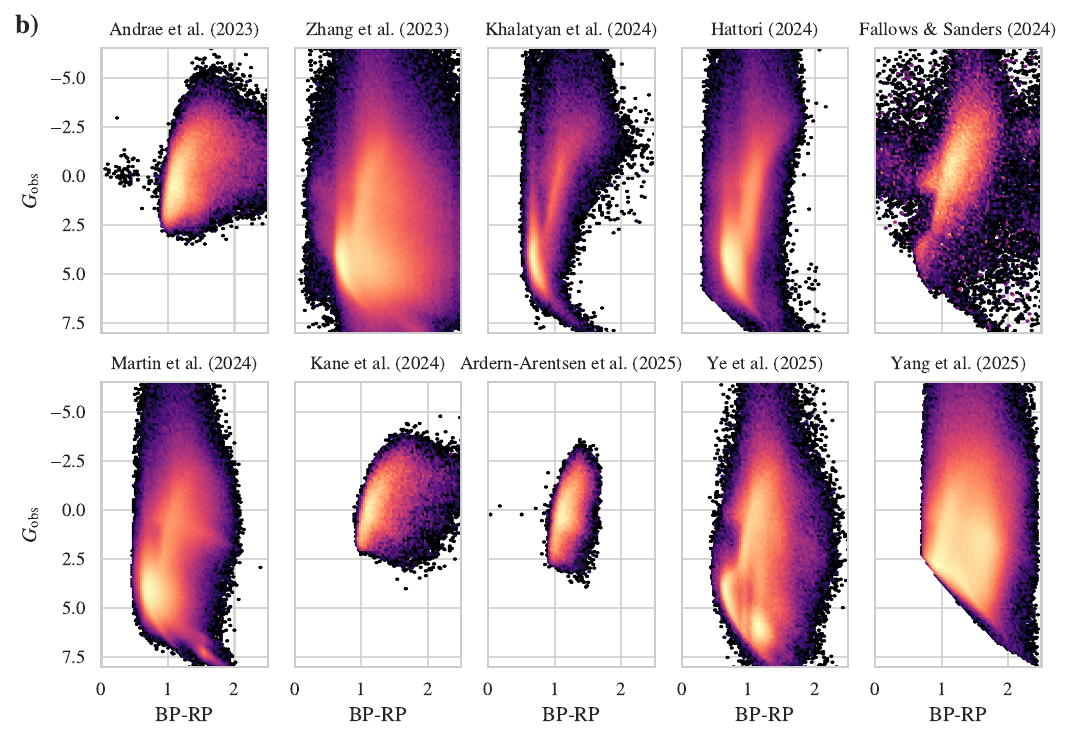}
\caption{Stellar density distributions in the Gaia CMD. Panels in (a) display densities based on apparent magnitude, while panels in (b) show densities using observed absolute magnitudes derived from Gaia parallaxes. Data are sourced from various Gaia XP metallicity catalogs following their quality cuts listed in Table \ref{tab:catalog_summary}. The quality cuts restrict several catalogs to mostly giants due to training pipelines and output quality control.  }
\label{fig:gaia_xp_cmd}
\end{figure*}

\subsection{Zero Rotation Expected for the Proto-Galaxy}
Figure \ref{fig:gaia_xp_cmd} shows the stellar density distribution in the Gaia color-magnitude diagram (CMD) for various Gaia XP spectroscopic metallicity catalogs. Panels (a) use apparent magnitudes $G$ and BP-RP colors, while the panels (b) use observed absolute magnitudes $G_\mathrm{obs}$ derived from parallax without extinction or reddening corrections.
\begin{equation}
  G_{\mathrm{obs}} \equiv G + 5 \log_{10}\!\bigl(\varpi_{\mathrm{mas}}\bigr) - 10,
\end{equation}
Panels a) provide a visual representation of the quality cuts per the authors' recommendations listed in the last column of Table \ref{tab:catalog_summary}, while panels (b) reduce distance-driven smearing and makes the CMDs more comparable across catalogs, at the cost of relying on parallax quality. {We do not apply an additional dereddening correction because the retained samples have already been filtered into relatively low- or moderate-extinction regimes by the adopted catalog-specific quality cuts ($E(B-V) < 0.1$--$0.5$), and several pipelines explicitly account for extinction or operate on corrected XP spectra (see Table~\ref{tab:catalog_summary}).} 

XP spectra exist only for sources with $G<17.65$. In addition, many catalogs apply explicit or implicit cuts that confine their inference domains to their training samples (e.g., RGB/RC giants or specific $T_{\mathrm{eff}}$-$log\,g$ ranges). These choices produce visible differences in CMD occupancy across catalogs: giant-only catalogs emphasize the upper RGB and red clump stars, whereas catalogs trained on mixed luminosity classes populate the main-sequence turnoff (MSTO) and subgiant branch as well. The footprints of the catalogs on the CMD could potentially bias the metallicity distribution in two ways: 1) At fixed age and evolutionary phase, a more metal-poor star would be brighter; 2) the magnitude limits changes the spatial coverage and the mixture of inner vs. outer halo stars. 

{The distinct CMD footprints of these catalogs, resulting from their different training sets and quality cuts, imply different stellar selection functions. A catalog restricted to bright giants will probe a larger volume and thus a different mix of Galactic components compared to a catalog dominated by main-sequence dwarfs in the solar neighborhood. Furthermore, pipelines may have varying completeness and contamination rates; for instance, a model trained primarily on red giants might misclassify metal-rich dwarfs as metal-poor if their colors are similar, or fail to recover genuine metal-poor stars in sparsely sampled regions of the CMD. These effects can alter the apparent normalization and shape of the MDF by changing the relative contributions of different Galactic components and by introducing metallicity-dependent systematic errors.}

\begin{figure*}[pht!]
\epsscale{0.9}
\plotone{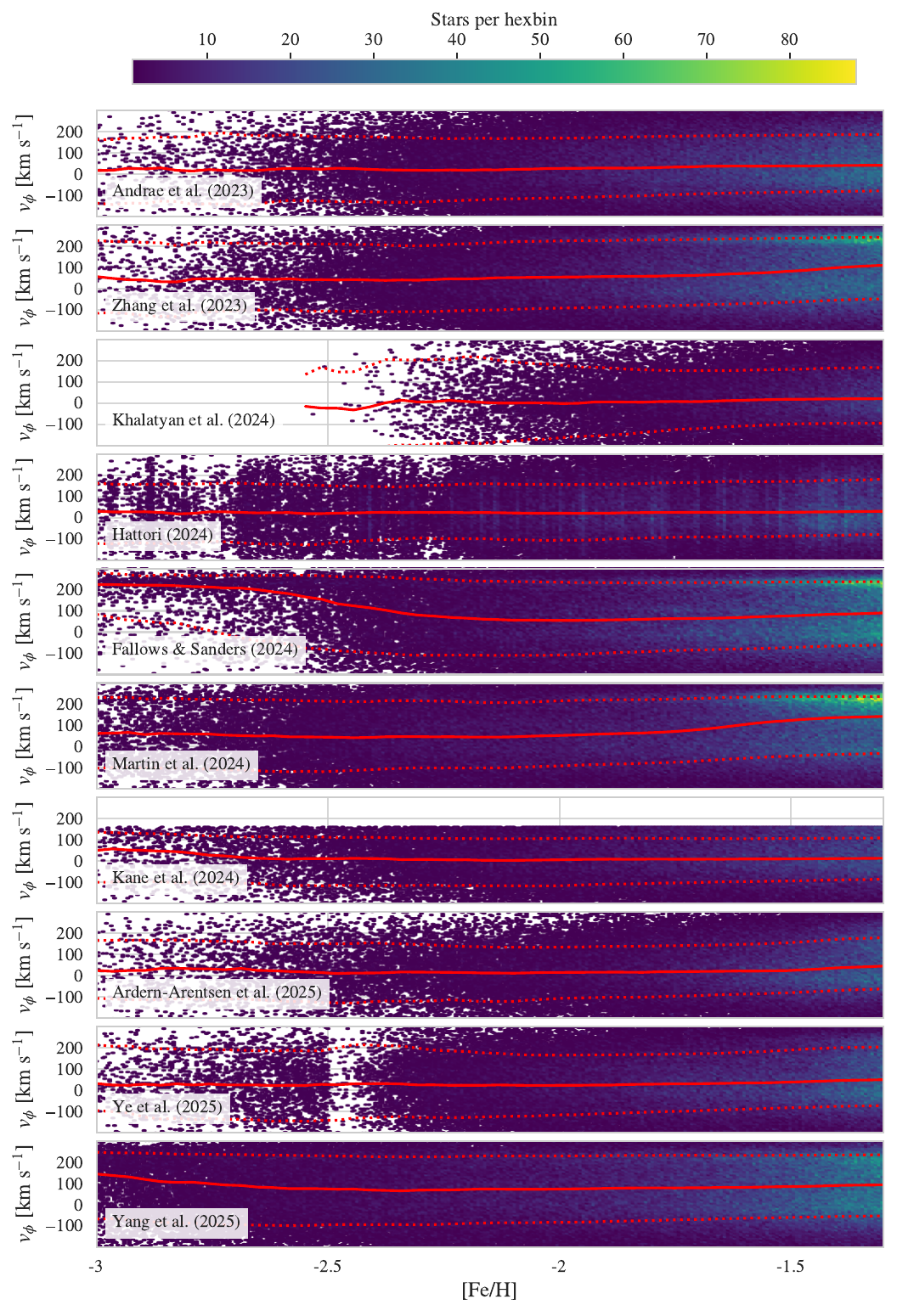}
\caption{Stellar distribution in Galactocentric rotational velocity versus metallicity for various Gaia XP metallicity catalogs. The solid red curve shows the running median $v_\phi$ as a function of metallicity, computed in a sliding window of width 0.2~dex in $\mathrm{[Fe/H]}$. The dotted lines show the 16th and 84th percentile range of $v_\phi$ respectively within the same window. Most metal-poor stars have low rotational velocities as expected for the proto-Galaxy, but false metal-poor stars that likely belong to the disk become common as [Fe/H] exceeds -2.0. Similar contamination also appears in \citet{2024MNRAS.531.2126F} and \citet{2025ApJS..279....7Y} as [Fe/H] drops below -2.6. }
\label{fig:gaia_xp_feh_vphi}
\end{figure*}

\begin{figure*}[pht!]
\epsscale{0.9}
\plotone{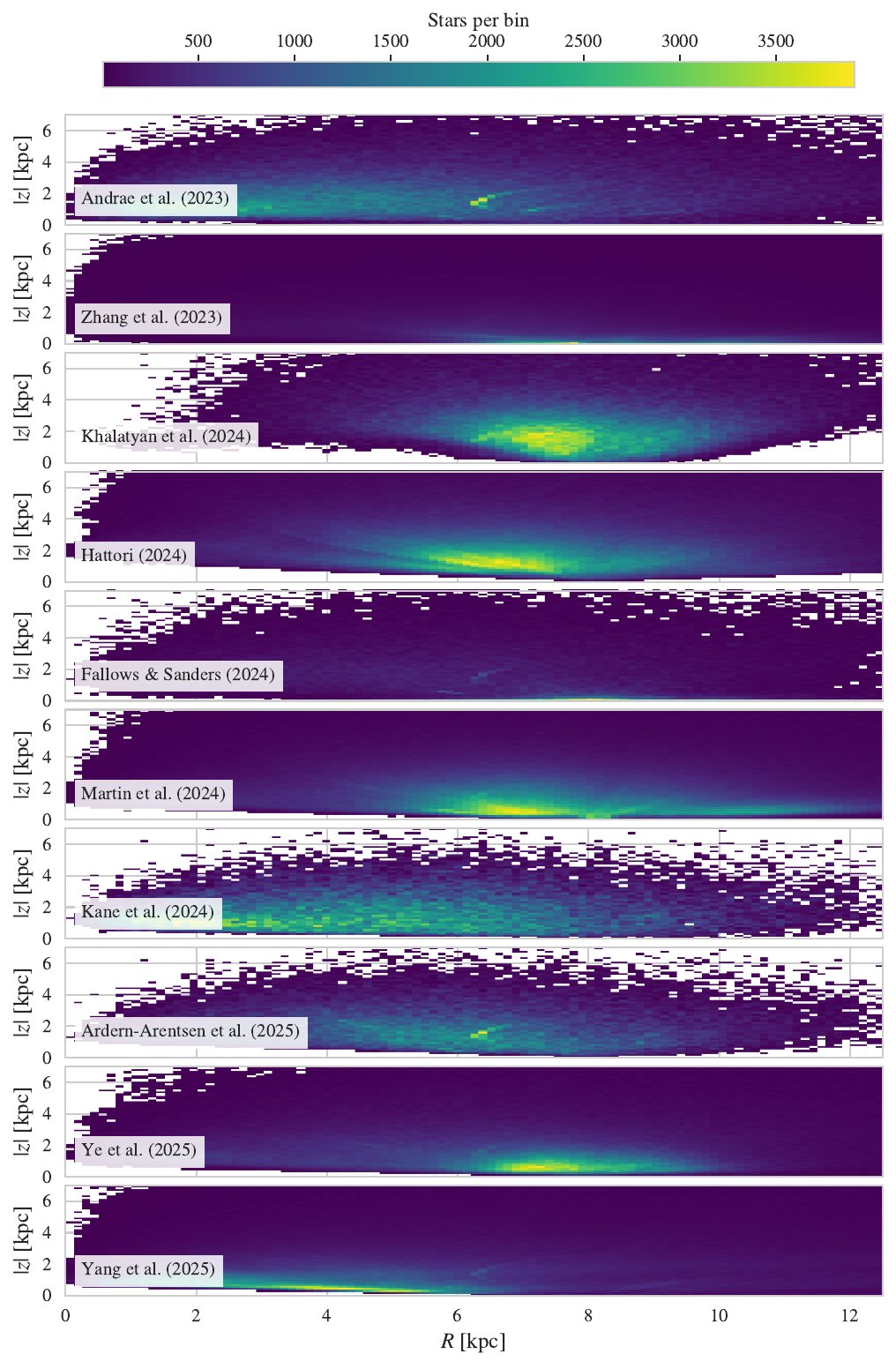}
\caption{Stellar distribution in Galactocentric radius ($R$) versus height above the Galactic plane ($|\mathrm{z}|$) for the Gaia XP metallicity catalogs. Differences in training sets and quality controls produce distinct spatial footprints. Catalogs restricted to giants extend to larger $R$ and $|\mathrm{z}|$, whereas others more closely follow the native XP selection function as being concentrated in the solar neighborhood. A notable outcome of Gaia XP is the recovery of large numbers of metal-poor stars at low $|\mathrm{z}|$, an area that was sparsely sampled before Gaia XP.}
\label{fig:gaia_xp_R_z}
\end{figure*}

\begin{figure*}[ht!]
\plotone{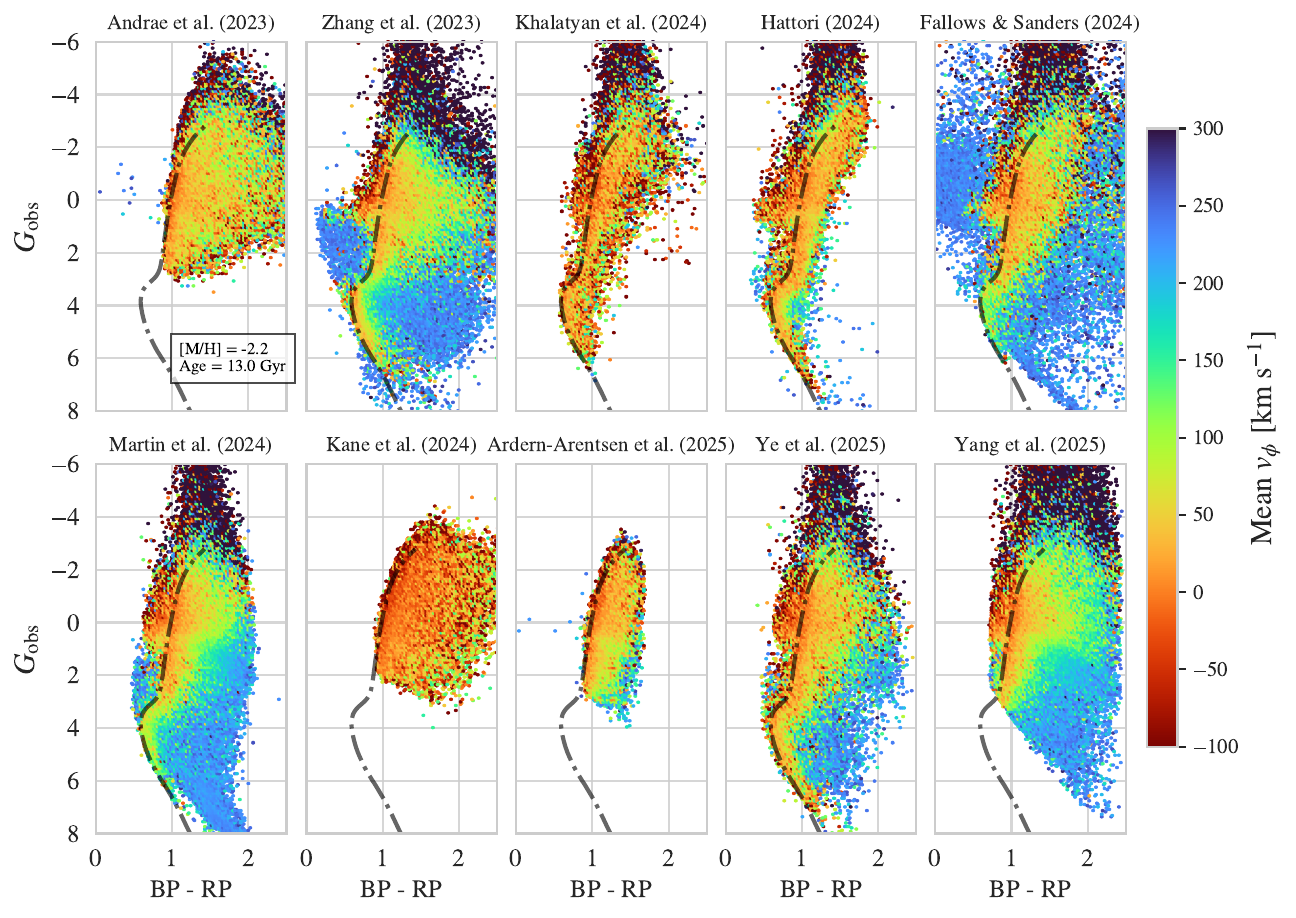}
\caption{Gaia CMD colored by median Galactocentric rotational velocity ($v_\phi$) in various Gaia XP metallicity catalogs. The magnitudes are observed absolute magnitudes as in b) in Fig.~\ref{fig:gaia_xp_cmd}. The slow-rotating stars highlight a PARSEC 1.2S isochrone \citep{2012MNRAS.427..127B} for a stellar population 13 Gyr old and with [M/H] = -2.2. Most stars with radial velocities in the various catalogs also tend to reside along such an isochrone on the CMD as seen in Figure \ref{fig:gaia_xp_cmd}. This suggests that at least most stars with low $v_\phi$ in Gaia RVS are genuinely old metal-poor stars.     }
\label{fig:gaia_xp_hr_vphi}
\end{figure*}

{To assess possible contamination, we derived Galactocentric positions and velocities only for the subset of stars with published Gaia DR3 RVS radial velocities, which constitutes approximately 5--10\% of each Gaia XP metallicity catalog after the adopted quality cuts. These kinematic data are used solely as an external diagnostic of the metal-poor population and were not used to define the MDF samples or to impose kinematic cuts on the full catalogs.} Stars with [Fe/H]$< -1.5$ are expected to have near-zero mean rotation with few objects reaching disk-like azimuthal velocities $v_\phi \gtrsim 100~\mathrm{km\,s^{-1}}$ \citep{2000AJ....119.2843C, 2007Natur.450.1020C, 2011MNRAS.411.1480D, 2018ApJ...856L..26M, 2024A&A...691L...1Y}. To test this expectation and diagnose catalog-dependent artifacts, we {transformed the Gaia astrometry together with the available Gaia DR3 RVS radial velocities for this subset into Galactocentric coordinates} ($R$, $\phi$, $z$) and velocities ($v_r$, $v_\phi$, $v_z$) using \texttt{astropy} \citep{2022ApJ...935..167A}. {Because Gaia RVS radial velocities are available for only a minority of XP sources, these kinematic distributions should be interpreted as a validation subset rather than as a complete description of the parent metallicity catalogs.}

Figure \ref{fig:gaia_xp_feh_vphi} presents the two-dimensional stellar density distribution of $v_\phi$ versus [Fe/H] for $-3 \le [\mathrm{Fe/H}]< -1.5$ in the metallicity catalogs. The running median $v_\phi$ is marked in each panel with a red solid line and the 16th and 84th percentile with dotted lines. The catalog by \citet{2024arXiv240900197K} applied a cut of $v_\phi < 160$ $\mathrm{km\,s^{-1}}$. In several catalogs, we observe a concentration of stars with $v_\phi \approx 220$ $\mathrm{km\,s^{-1}}$ as [Fe/H] approaches our boundary (typically $-1.8 \lesssim [\mathrm{Fe/H}] < -1.5$), most prominently in \citet{2023MNRAS.524.1855Z, 2024MNRAS.531.2126F, 2024A&A...692A.115M, 2025ApJS..279....7Y}. At the very metal-poor end with [Fe/H] $< -2.5$, the median $v_\phi$ differs across catalogs: in \citet{2024MNRAS.531.2126F, 2025A&A...695A..75Y, 2025ApJS..279....7Y} it exceeds $50~\mathrm{km\,s^{-1}}$, while in the remaining catalogs it stays closer to $0$. The high-$v_\phi$ features near the metallicity boundaries can arise from: (1) misclassification of metal-rich disk stars as metal-poor; (2) genuine metal-poor stars on fast, disk-like orbits associated with the metal-weak thick disk or ancient, rotation-dominated components \citep{2021MNRAS.508.1509F, 2024A&A...683A.136B, 2024A&A...691L...1Y, 2025NatAs...9..101X}.   

A physically motivated channel for genuinely fast, metal-poor stars could be an early, gas-rich merger that seeded the proto-disk's angular-momentum axis, producing a prograde, low-$\mathrm{[Fe/H]}$ tail \citep{2021MNRAS.505..921S}. An alternative is that later mergers, such as the GSE, could have torqued pre-existing stellar populations, imparting net rotation \citep{2024MNRAS.527.6926M}. Whether the observed prograde motions are primordial or merger-induced, their kinematic coherence suggests a link to the early building blocks of the proto-Galaxy \citep{2019MNRAS.484.2166S, 2021ApJ...908..208C}. However, as shown in Fig.~\ref{fig:gaia_xp_feh_vphi}, the bulk of stars with $\mathrm{[Fe/H]} < -2.0$ exhibit low $v_\phi$, consistent with a dispersion-dominated proto-Galaxy. Because only $\sim 5$--10\% of stars in each catalog have Gaia RVS radial velocities, kinematic cuts alone cannot be applied to the entire catalogs. An alternative approach is to apply isochrone-based selections, which can be done with the available photometry and helps mitigate contamination. We restrict our MDF analysis to $ \mathrm{[Fe/H]} < -2.0$ to minimize the inclusion of disk stars. Since most stars already follow the observed isochrone for an old, metal-poor stellar population and most RVS stars have low rotation, we leave a detailed kinematic decomposition for future work.

Figure~\ref{fig:gaia_xp_R_z} shows the two-dimensional number density in Galactocentric radius $R$ and height $|\mathrm{z}|$ for the XP catalogs we used within our metallicity range. Unlike many previous studies that targeted high-$|\mathrm{z}|$ regions, our working samples are overwhelmingly dominated by stars in the disk plane: most stars lie at $|\mathrm{z}|<2$ kpc because the underlying Gaia XP selections are magnitude-limited and largely all-sky. The vetted RGB catalog by \citet{2023ApJS..267....8A} and the catalogs by \citet{2025MNRAS.536.2507K} and \citet{2025ApJS..279....7Y} target giants, which results in more even $R$ coverage but naturally misses the dwarfs at low-$|\mathrm{z}|$. The pipeline by \citet{2023ApJS..267....8A} require WISE $W1/W2$ and even use $M_{W1}$ and $G-W$ cuts to purify giants. CatWISE was adopted precisely to improve completeness over AllWISE\footnote{\url{https://irsa.ipac.caltech.edu/data/WISE/docs/release/AllWISE/}} in crowded fields. Consequently, some residual suppression in the dustiest mid-plane lines of sight is expected.

\citet{2024MNRAS.530.3391A} applied their CEMP-focused network to the same vetted RGB sample so their $R$--$|\mathrm{z}|$ distribution follows the same footprint. \citet{2023MNRAS.524.1855Z} assumed a single extinction law and applied it across the full XP sample, which yields the strongest mid-plane concentration at a given $R$ because it largely follows the raw Gaia density. Similarly, \citet{2024MNRAS.531.2126F} included broad-band (2MASS/WISE) colors with XP and publish an all-sky catalog the mirrors XP density. SHBoost by \citet{2024A&A...691A..98K} was applied to all Gaia XP sources but their quality flags assign the highest metallicity reliability to stars with low extinction. \citet{2025ApJ...980...90H} restricts to low-extinction sightlines, which removes the opaque mid-plane and accentuates higher-$|\mathrm{z}|$ at fixed $R$. The Pristine-Gaia synthetic catalog by \citet{2024A&A...692A.115M} iteratively corrects for extinction and was explicitly built to cover all Galactic components on the sky, which yields good plane coverage with only modest suppression in the dustiest lines of sight. \citet{2025A&A...695A..75Y} cautioned that reliability for their metallicity measurements is the strongest for roughly $4000\lesssim T_{\rm eff}\lesssim7000$ K, effectively emphasizing the FGK-rich mid-plane region in practice. 

Figure~\ref{fig:gaia_xp_hr_vphi} maps median $v_\phi$ across the CMD (using $G_{\mathrm{obs}}$), localizing, via the colour scale spanning from $-100$ to $300~\mathrm{km\ s^{-1}}$, where prograde stars concentrate in evolutionary phase. Stars with median $|v_\phi| < 50~\mathrm{km\ s^{-1}}$ trace a PARSEC 1.2S isochrone for an old, metal-poor population ($\mathrm{[M/H]}=-2.2$, $\mathrm{Age}=13\,\mathrm{Gyr}$), as expected, while stars that deviate from the isochrone—typically toward the blue turnoff and red subgiant/dwarf edge—show elevated $v_\phi$ over $200~\mathrm{km\ s^{-1}}$. These concentrations with high $v_\phi$ on the CMD coincide with regions where training set coverage is sparse because more than half the catalogs are trained on or anchored to APOGEE. For example, \citet{2024MNRAS.531.2126F} trained their NNs on APOGEE data and \citet{2025A&A...695A..75Y} used APOGEE to train their NNs for flux correction. Even broader ``transfer'' models like SHBoost by \citet{2024A&A...691A..98K} still draw much of their label scale from spectroscopic surveys that under-sample parts of the CMD and thus inherit the coverage gaps. Therefore, the high-$v_\phi$ patches are most naturally interpreted as thin-disk contaminants whose metallicities are severely underestimated by the pipelines. 

A similar trend is seen in independent photometric-metallicity work by \citet{2024A&A...692A.115M}: CaHK-syn metallicity estimates degrade toward hotter stars and lose sensitivity for dwarfs. Substituting their default (giant-calibrated) metallicities with the companion dwarf-model values did not alter the predominance of high-$v_\phi$ dwarfs. The difference in [Fe/H] between the two models is generally 0.2--0.4 dex, which is small compared to our metallicity range. Catalogs that confine their labels to well-calibrated regions of the CMD generally avoid spurious high-$v_\phi$ patches. For example, \citet{2023ApJS..267....8A} released a vetted RGB-only catalog. \citet{2025MNRAS.537.1984A} also selected their objects partially from this RGB sample. More broadly, a training-set-dependent CMD mask-such as the flag provided by \citet{2025ApJ...980...90H}-offers a practical way to refine samples and minimize contamination.

\subsection{Robust MDF Slope Across Catalogs and Selections }

\begin{figure*}[ht!]
\plotone{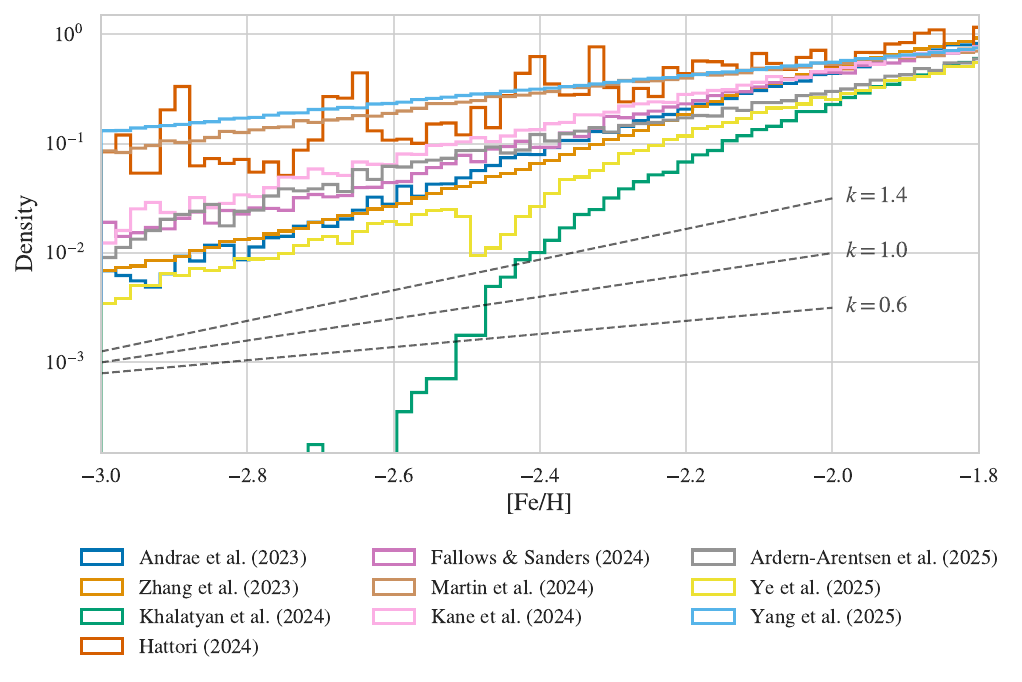}
\caption{Logarithmic metal-poor MDFs from various Gaia XP catalogs after applying the quality cuts listed in Table~\ref{tab:catalog_summary}. {Bin width is for this and the panels in the following two figures is 0.02 dex.} Dashed lines show exponential density profiles with reference slopes $k = 0.6$, 1.0, and 1.4 for comparison. Despite differences among the catalogs, most exhibit a near-exponential profile with $k \approx 1.0$. Catalogs from \citet{2024A&A...691A..98K} and \citet{2025A&A...695A..75Y}, trained solely on APOGEE, under-represent VMP stars. {Although this and the following figures could extend to $\mathrm{[Fe/H]}>-2.0$, all slope estimates are restricted to $-3 < \mathrm{[Fe/H]} < -2.0$.}}
\label{fig:gaia_xp_mdf}
\end{figure*}

\begin{figure*}[ht!]
\plotone{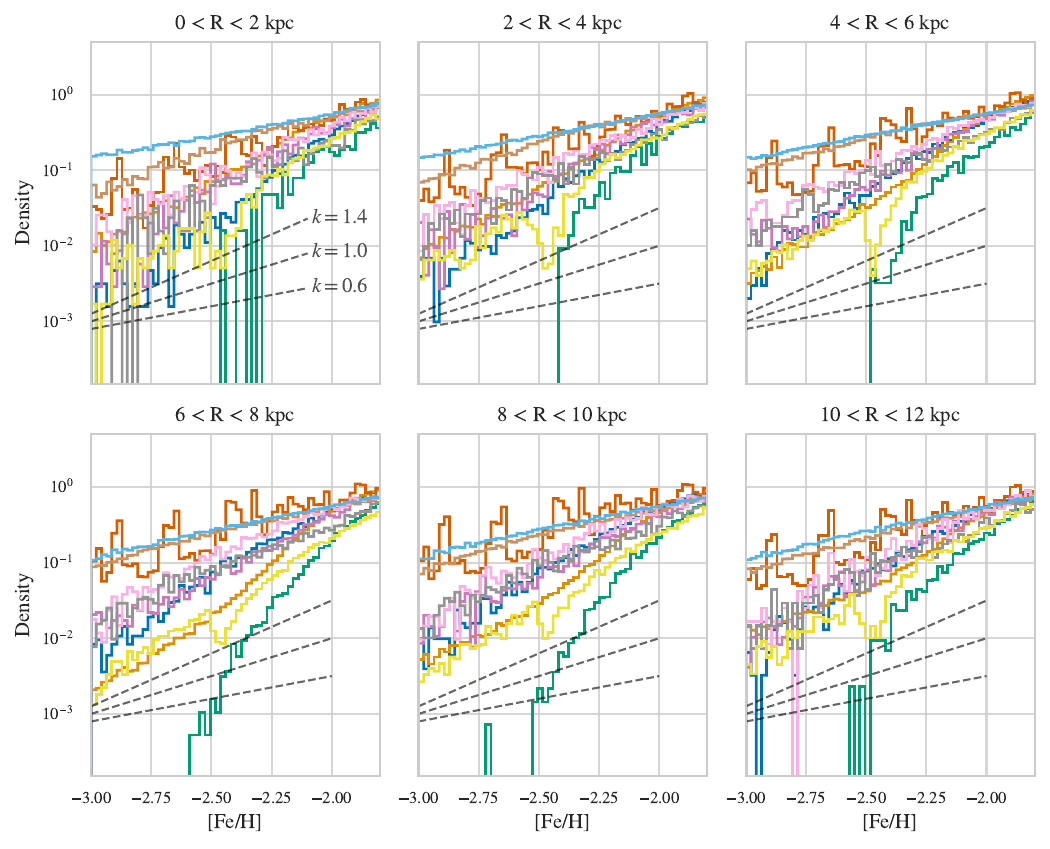}
\epsscale{0.6}
\caption{Logarithmic metal-poor MDFs from Gaia XP metallicity catalogs in bins of Galactocentric radius $R$. The panels show MDFs in 2 kpc intervals from $0$ to $12$ kpc. In each panel, dashed lines indicate exponential density profiles with three reference slopes ($k = 0.6, 1.0, 1.4$) for comparison, as in Fig.~\ref{fig:gaia_xp_mdf}. The inferred MDF slopes remain broadly consistent across radial bins, except where the number of stars is too small to define the MDF robustly.}
\label{fig:gaia_xp_mdf_R}
\end{figure*}

\begin{figure*}[ht!]
\plotone{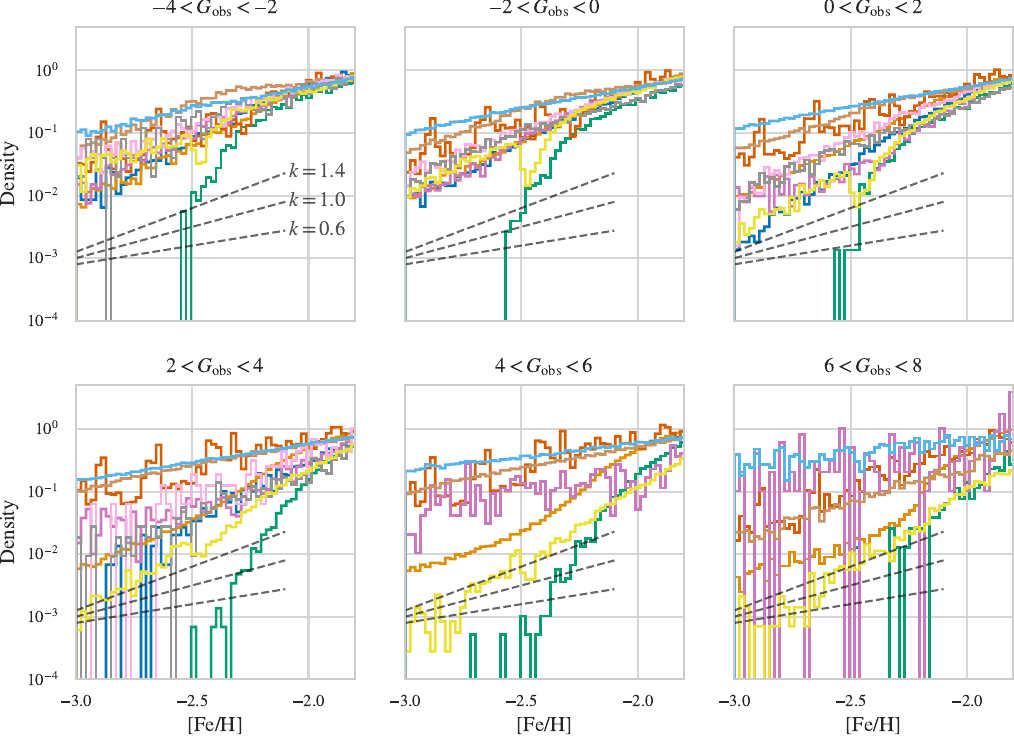}
\epsscale{0.6}
\caption{Logarithmic metal-poor MDFs from Gaia XP metallicity catalogs in bins of parallax-corrected magnitude, $G_\mathrm{obs}$. The panels show MDFs for subsamples divided by $G_\mathrm{obs}$ estimated from the observed $G$ magnitude and parallax, without correcting for extinction. In each panel, dashed lines indicate exponential density profiles with three reference slopes ($k = 0.6, 1.0, 1.4$) for comparison, as in Fig.~\ref{fig:gaia_xp_mdf}. The inferred MDF slopes remain broadly consistent across $G_\mathrm{obs}$ bins, except where the number of stars is too small to define the MDF robustly.}
\label{fig:gaia_xp_mdf_Gobs}
\end{figure*}

{To measure the slope of the metal-poor tail, we fit each catalog over the fixed interval $-3 \le \mathrm{[Fe/H]} \le -2$. Rather than fitting a straight line to binned histograms, we use the individual stellar metallicities directly. This avoids sensitivity to bin size and reduces the impact of noisy low-count bins.}

{We model the MDF tail as an exponential function,
\begin{equation}
p(x) \propto \exp(\beta x), \qquad x \equiv \mathrm{[Fe/H]},
\end{equation}
which is equivalent to a straight line in logarithmic space,
\begin{equation}
\log_{10} p(x) = kx + c, \qquad k = \beta / \ln 10.
\end{equation}
We determine $k$ with a maximum-likelihood fit using only stars in the chosen metallicity range. To estimate the uncertainty, we use bootstrap resampling. For each catalog, we repeatedly resample the stars with replacement, refit the slope, and take the median and 16th--84th percentile range of the resulting slope distribution. We report these as the best-fit value and uncertainty for each catalog in Table \ref{tab:mdf_slopes_truncexp}.}

\begin{table}[ht!]
\centering
\caption{MDF tail slopes estimated over $-3 \le \mathrm{[Fe/H]} \le -2$ using an unbinned maximum-likelihood fit to a truncated exponential model. Reported values are the bootstrap median and 16th--84th percentile interval.}
\label{tab:mdf_slopes_truncexp}
\begin{tabular}{c l c}
\hline
Color & Catalog & $k$ \\
\hline
\textcolor[HTML]{0173B2}{\rule{1.2ex}{1.2ex}} & Andrae et al. (2023) & $1.93^{+0.01}_{-0.01}$ \\
\textcolor[HTML]{DE8F05}{\rule{1.2ex}{1.2ex}} & Zhang et al. (2023) & $2.09^{+0.00}_{-0.00}$ \\
\textcolor[HTML]{029E73}{\rule{1.2ex}{1.2ex}} & Khalatyan et al. (2024) & $3.44^{+0.03}_{-0.03}$ \\
\textcolor[HTML]{D55E00}{\rule{1.2ex}{1.2ex}} & Hattori (2024) & $0.92^{+0.00}_{-0.00}$ \\
\textcolor[HTML]{CC78BC}{\rule{1.2ex}{1.2ex}} & Fallows \& Sanders (2024) & $1.65^{+0.01}_{-0.01}$ \\
\textcolor[HTML]{CA9161}{\rule{1.2ex}{1.2ex}} & Martin et al. (2024) & $0.79^{+0.00}_{-0.00}$ \\
\textcolor[HTML]{FBAFE4}{\rule{1.2ex}{1.2ex}} & Kane et al. (2024) & $1.40^{+0.02}_{-0.02}$ \\
\textcolor[HTML]{949494}{\rule{1.2ex}{1.2ex}} & Ardern-Arentsen et al. (2025) & $1.26^{+0.02}_{-0.02}$ \\
\textcolor[HTML]{ECE133}{\rule{1.2ex}{1.2ex}} & Ye et al. (2025) & $2.07^{+0.01}_{-0.01}$ \\
\textcolor[HTML]{56B4E9}{\rule{1.2ex}{1.2ex}} & Yang et al. (2025) & $0.63^{+0.00}_{-0.00}$ \\
\hline
\end{tabular}
\end{table}

Figure \ref{fig:gaia_xp_mdf} shows the MDFs over $-3<\mathrm{[Fe/H]}<-1.8$ on a logarithmic scale for the Gaia XP catalogs listed in Sec.~\ref{sec:data}. Using log density highlights the near-exponential falloff of the metal-poor tail, as in Fig.~4 of \citet{2022ApJ...941...45R} for $-2<\mathrm{[Fe/H]}<-1$. For visual guidance, dashed lines indicate three reference slopes ($k=0.6, 1.0, 1.4$) that bracket the range we observe across catalogs in this interval. We excluded \citet{2024ApJS..272....2L}, whose metallicities are truncated at $\mathrm{[Fe/H]}=-2.0$, producing an artificial pile-up at the boundary. 

Because \citet{2024A&A...691A..98K} and \citet{2025A&A...695A..75Y} are anchored to APOGEE training labels, their MDFs decline rapidly below $\mathrm{[Fe/H]}\approx-2$, where APOGEE offers few measurements. In addition, \citet{2025A&A...695A..75Y} disable their APOGEE-based NN flux correction for stars with $\mathrm{[Fe/H]}\le-2.5$ and fit \texttt{FERRE} directly, which they release as a separate metal-poor subset. This treatment imprints a discontinuity near $\mathrm{[Fe/H]} = -2.5$, but the MDF tail slope remains consistent. \citet{2025ApJS..279....7Y} train against metal-poor reference sets (PASTEL, SAGA) with a cost-sensitive objective to enhance recovery of VMP giants, similarly flattening their tail. Finally, \citet{2024A&A...692A.115M}-as part of the Pristine effort targeting $\mathrm{[Fe/H]}<-1$-deliver a comparatively larger fraction of metal-poor stars from synthetic CaHK, which also yields a flat density decline in their MDF. 

The similarly flat MDF tail in the \citet{2025ApJ...980...90H} catalog is best explained by two design choices that boost recovery of VMP halo stars: (1) the training set was deliberately augmented with the VMP sample of \citet{2022ApJ...931..147L}, extending label coverage well below $\mathrm{[M/H]}\approx-2.5$ and increasing counts at the low-metallicity end; and (2) the recommended catalog is restricted to low-extinction sight lines, $E(B-V)<0.1$, which preferentially probe high-latitude halo volume and improves completeness for distant halo giants. However, the same VMP set was also used for training by \citet{2023MNRAS.524.1855Z} and \citet{2025MNRAS.537.1984A}, yet both deliver noticeably steeper metal-poor tails. This suggests that simply augmenting training set with VMP stars does not necessarily determine the slope; sample selection and quality cuts play a more important role. 

\citet{2023ApJS..267....8A} and \citet{2023MNRAS.524.1855Z} appear to under-represent the most metal-poor regime: their MDFs show the steepest slope for $\mathrm{[Fe/H]}<-2.0$ with $k > 1.4$, then flatten toward $k \approx 1.0$ as metallicity increases. \citet{2023ApJS..267....8A} selected sources in their vetted RGB sample to avoid spurious metal-poor stars, which could preferentially remove some VMP stars. \citet{2023MNRAS.524.1855Z} did not explicitly include VMP stars in their training set and their basic reliability cuts tend to cull low-SNR sources, disproportionately trimming the VMP tail. Overall, the metal-poor tails have logarithmic slopes centered near $k \approx 1$, and are well described by a single exponential over our metallicity window. We observe no bumps or peaks within $-3<\mathrm{[Fe/H]}<-1.8$ after applying recommended quality cuts from the authors. Minor departures from linearity in log-density appear only at the boundaries-flattening in VMP-boosted catalogs and steepening where training or completeness drops below $\mathrm{[Fe/H]} \approx -2.0$-and are consistent with known selection and pipeline effects.

{Figure~\ref{fig:gaia_xp_mdf_R} shows MDFs over $-3<\mathrm{[Fe/H]}<-1.8$ split by Galactocentric radius $R$, while Figure~\ref{fig:gaia_xp_mdf_Gobs} shows the same MDFs split by parallax-corrected magnitude, with the same reference slopes from Fig.~\ref{fig:gaia_xp_mdf}. Because the $y$-axis is logarithmic, the curves tend to meet near $\mathrm{[Fe/H]}\simeq-1.8$ at density of 1. A quick way to read an individual slope is to note the intercept at $\mathrm{[Fe/H]}=-3$: for example, an intercept of $10^{-1}$ implies $k\approx 1.0/1.5\simeq0.67$, while $10^{-2}$ implies $k\simeq1.33$. Most catalogs exhibit nearly identical slopes across the various $R$ and parallax-corrected magnitude bins. The consistency across radius indicates that the MDF slope is not strongly sensitive to distance from the Galactic center, while the consistency across parallax-corrected magnitude suggests that the result is robust across different stellar types and luminosity classes despite the differing CMD footprints of the catalogs.}

{In Fig.~\ref{fig:gaia_xp_mdf_Gobs}, several samples contain fewer VMP stars at fainter parallax-corrected magnitudes. This is expected because some catalogs are dominated by giants and therefore contain relatively few intrinsically faint stars, while others have smaller overall samples in those bins. The MDF by \citet{2023MNRAS.524.1855Z} loses VMP stars more rapidly toward the faint end, consistent with their basic reliability cut, which preferentially culls lower-S/N sources with larger uncertainties. By contrast, the catalogs from \citet{2025ApJ...980...90H}, \citet{2025ApJS..279....7Y}, and \citet{2024A&A...692A.115M} show nearly constant slopes across the parallax-corrected magnitude range. The first catalog supplemented its training set with VMP stars and uses a CMD-based quality filter that retains MSTO and main-sequence stars, while the latter two benefit from synthetic spectra in their pipelines, which help mitigate the scarcity of VMP dwarfs in the training data.}

\subsection{MDF Slope Tells Early Assembly Tale}

We used the GCE model outlined in Sec.~\ref{sec:gce} from our first paper to explore the physical conditions behind different slopes of the metal-poor tail of MDF. The GCE models presented here are pre-selected in Paper I based on the following criteria: 1) [Fe/H] should reach $-1.3 \pm 0.05$ dex after one Gyr; 2) [Mg/Fe] should reach $0.26 \pm 0.05$ dex after one Gyr; 3) [Mg/Fe] increases above 0.3 dex with a minimum increase of 0.05 dex while [Fe/H] surpasses -0.9 dex in the next Gyr. 

{We assigned each GCE model to one of three slope categories, $k=0.6$, $1.0$, or $1.4$, by comparing its generated MDF to three reference MDFs representing these slope classes. For each model, we first computed a normalized metallicity histogram over the common $\mathrm{[Fe/H]}$ bins, and then evaluated a chi-square-like mismatch statistic with respect to each reference MDF,
\begin{equation}
\chi^2 = \sum_i \frac{(M_i - R_i)^2}{R_i + 10^{-10}},
\end{equation}
where $M_i$ is the model MDF in bin $i$ and $R_i$ is the corresponding reference MDF. Each model was assigned to the category that gave the smallest mismatch. We retained only models with a minimum mismatch below a threshold of $\chi^2 < 1.5$; models with larger deviations were discarded. This procedure therefore grouped models according to which reference MDF shape they most closely resembled, rather than by fitting each model slope independently.}

\begin{figure}[ht!]
\plotone{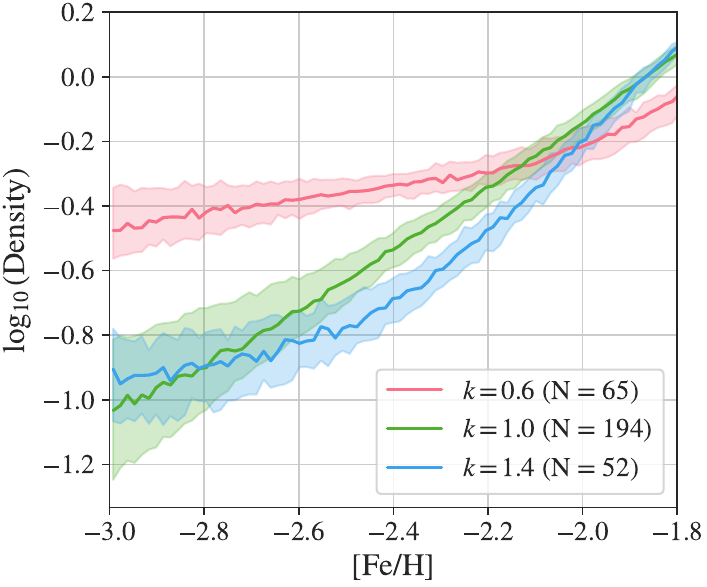}
\caption{Median MDFs from one-zone GCE models grouped by nearest reference MDF class corresponding to nominal slope $k=0.6,1.0,1.4$. The number of models within each slope category is annotated. Lines and shaded regions show the median and 16th--84th percentile ranges, respectively.   }
\label{fig:proto_mw_mdf}
\end{figure}


\begin{figure*}[pht!]
\epsscale{1.2}
\plotone{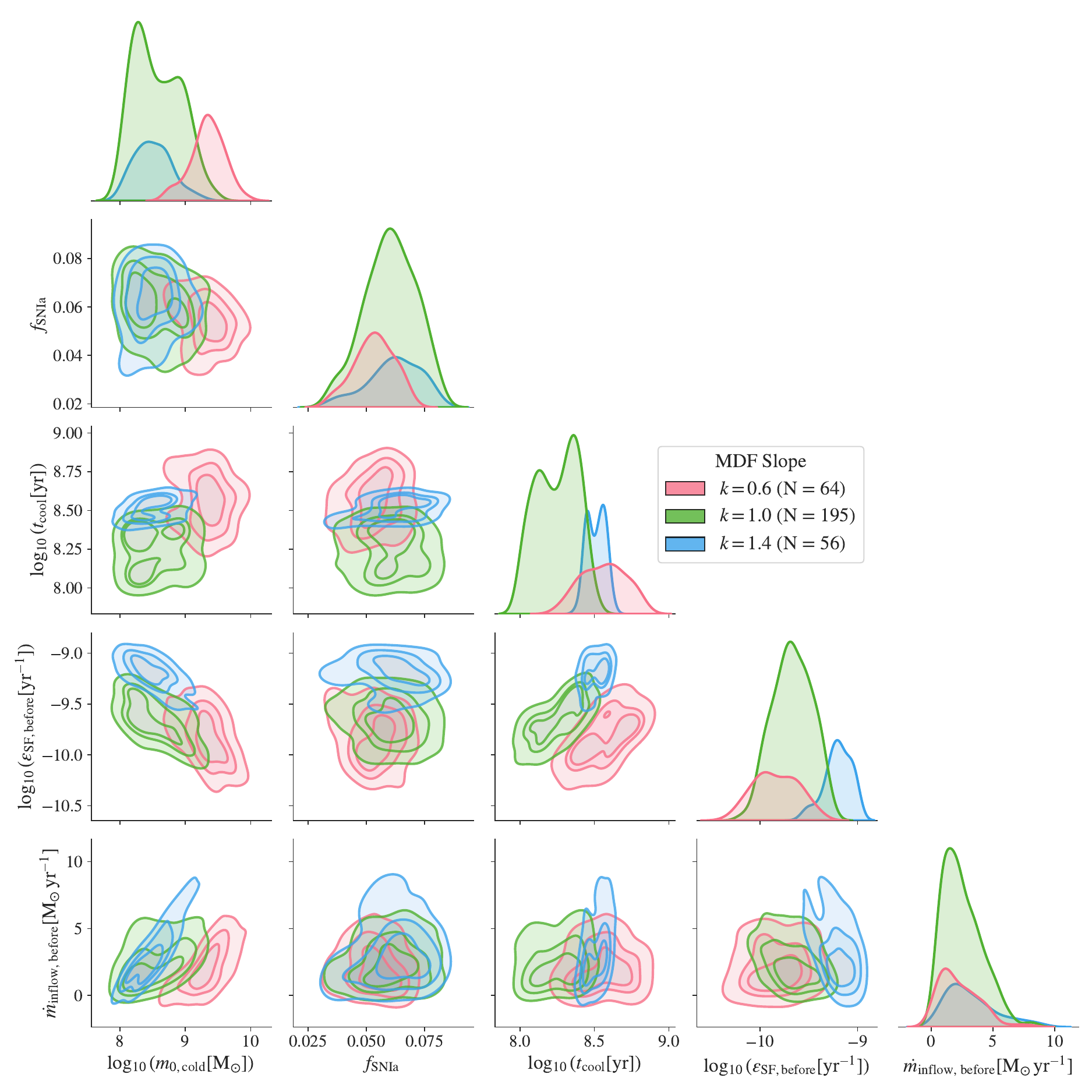}
\caption{
Two-dimensional marginal distributions of one-zone GCE model parameters, grouped by the metal-poor MDF exponential density slope $k$. The parameters, arranged left to right and bottom to top, are: the initial cold gas mass; the fraction of white dwarfs turning into SNe Ia; the warm ISM cooling timescale; the SFE during the initial proto-phase before the [$\alpha$/Fe] rise; and the pre-[$\alpha$/Fe]-rise gas inflow rate. One-dimensional marginal distributions are shown along the diagonal panels, while off-diagonal panels display kernel-density estimated contours enclosing 30\%, 60\%, and 90\% of the models. This plot highlights how variations in $k$ correlate with different GCE parameters.}
\label{fig:parameter_space}
\end{figure*}

\begin{figure}
\plotone{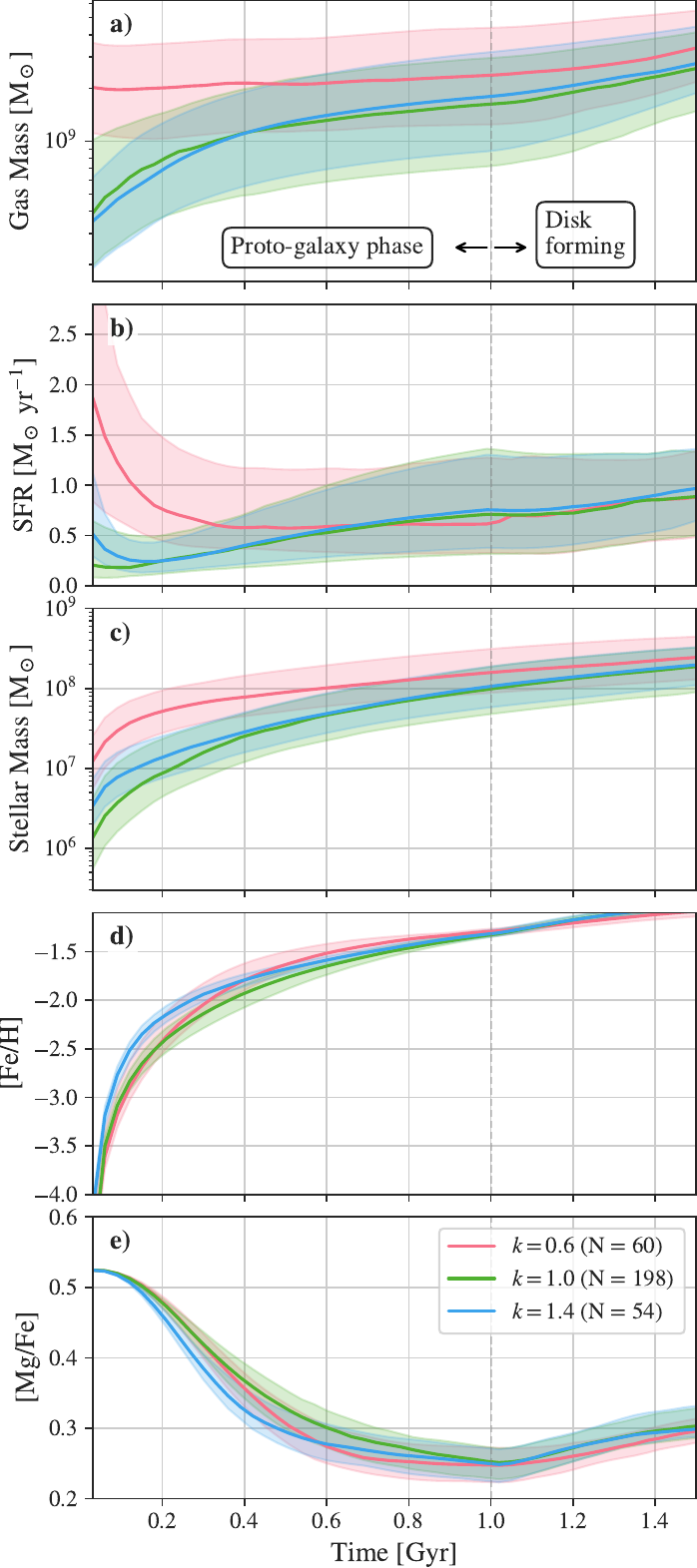}
\caption{Time evolution of global quantities in one-zone GCE models, grouped by the metal-poor MDF slope $k$. Panels show (a) total mass of cold and warm ISM, (b) star formation rate (SFR), (c) cumulative stellar mass surviving to the present, (d) stellar [Fe/H], and (e) stellar [Mg/Fe]. Solid curves indicate the median of models in each $k$ category, and shaded bands show the interquartile range (25-75th percentile). Models with $k = 0.6$ stand out with massive initial starbursts and near constant mass for their gas reservoirs, while models with $k = 1.4$ have moderate starbursts. }
\label{fig:proto_mw_gas_sfh_mdf}
\end{figure}

\begin{figure}[ht!]
\epsscale{1.0}
\plotone{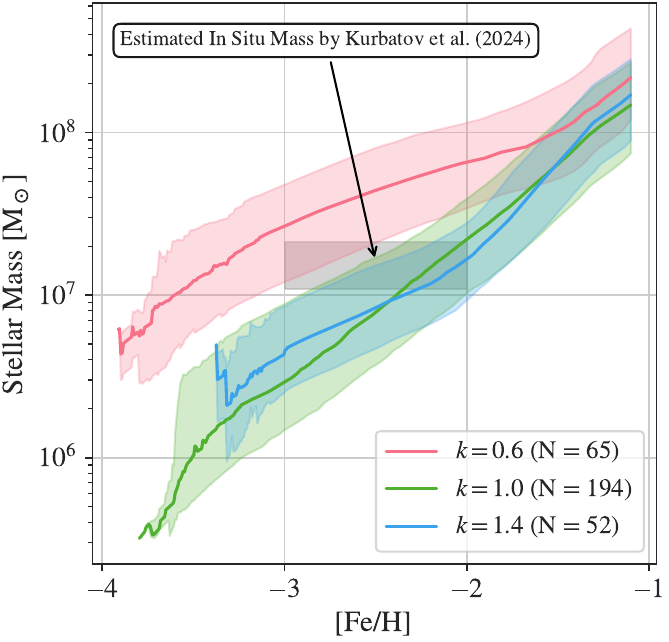}
\caption{The cumulative stellar mass assembly history as a function of metallicity ([Fe/H]) for three slope classes in our GCE models. Lines and shaded regions show the median and 16th--84th percentile ranges, respectively. \citet{2024arXiv241022250K} estimated the mass of the in situ proto-Galaxy to be between $1.07 \times 10^7$ and $2.11 \times 10^7\,\mathrm{M_\odot}$ between [Fe/H] = -3.0 and -2.0. The models with $k = 0.6$ can be ruled out with this restriction.}
\label{fig:cum_stellar_mass_vs_feh}
\end{figure}

\begin{figure}
\plotone{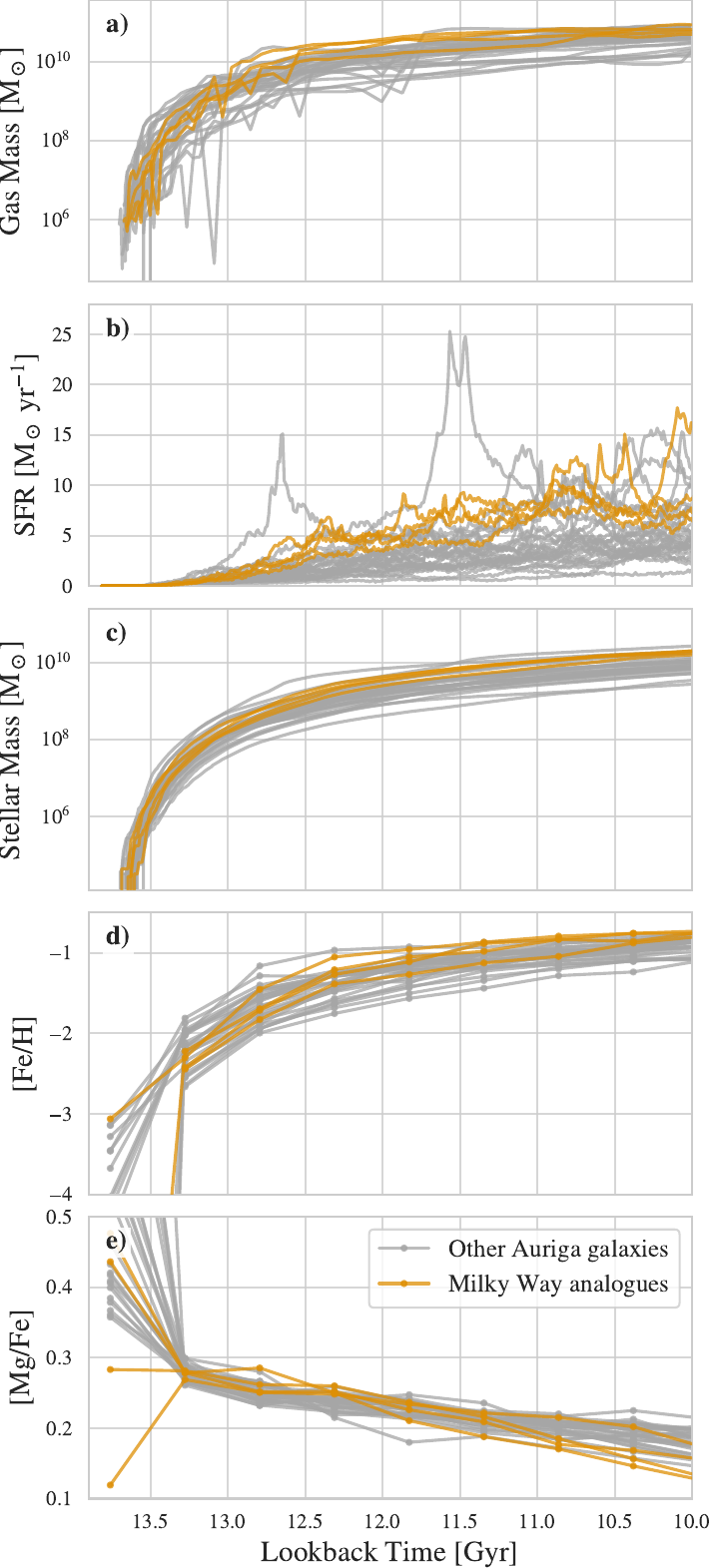}
\caption{Time evolution of global properties for Auriga galaxies. Panels show: (a) total gas mass; (b) SFR; (c) cumulative mass of stars surviving to $z=0$; (d) mass-weighted stellar [Fe/H]; and (e) mass-weighted stellar [Mg/Fe]. Tracks for the Milky Way analogues (Au-17, Au-18, Au-23, and Au-26) are highlighted in orange; the other Auriga galaxies are shown in gray. All of them show accumulation of gas and lack of starburst in the first Gyr, consistent with $k = 1.0$ models in Fig.\ref{fig:proto_mw_gas_sfh_mdf}.}
\label{fig:auriga_galaxy_global_properties}
\end{figure}

\begin{figure*}[ht!]
\plotone{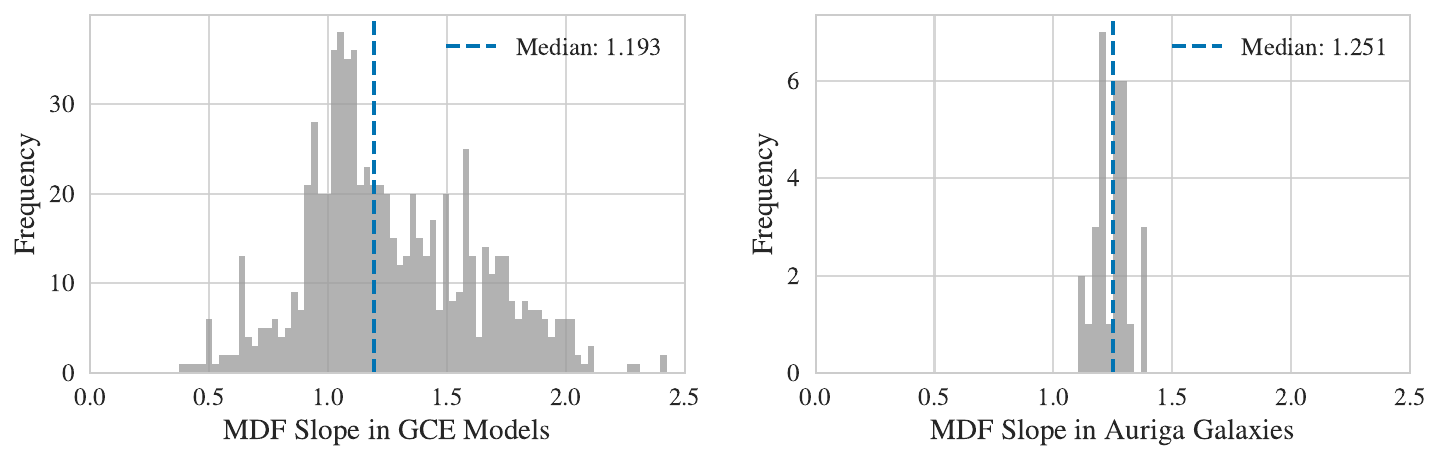}
\caption{Distribution of exponential density slopes fitted for the metal-poor MDFs from GCE models exhibiting an [$\alpha$/Fe]-rise and from Auriga galaxies measured between $-3.0 < \mathrm{[Fe/H]} < -2.0$. {The histograms are computed using a uniform bin width of 0.025 in slope.} The median slope for each group is marked with a vertical dashed line with its value annotated. The most common slopes among our GCE models with [$\alpha$/Fe]-rise agree with those produced in Auriga. The [$\alpha$/Fe]-rise is key to establishing the scenario of the proto-Galaxy and the observed MDF help us further narrow down.  }
\label{fig:Auriga_galaxy_hist_mdf_slopes}
\end{figure*}

Figure \ref{fig:proto_mw_mdf} shows the median MDFs of the subsets {of GCE models} that best match the three reference slopes from Fig.~\ref{fig:gaia_xp_mdf}. The shaded bands indicate the inter-model spread, and the count of contributing models is annotated beside each $k$ label. Except for $k=1.0$, a single exponential profile is an imperfect description: for $k=0.6$ the MDF steepens toward the metal-rich side once $\mathrm{[Fe/H]}\gtrsim-2.0$, whereas for $k=1.4$ it flattens at the metal-poor end below $\mathrm{[Fe/H]}\lesssim-2.4$. We fitted the MDFs piecewise with sliding windows of 0.5 dex over $-3.0 < \mathrm{[Fe/H]} < -1.5$. We then retained the windows whose slopes agreed and showed small residuals in log-density, and took their common value as the slope of the metal-poor tail of each MDF. The main limitation is imposed by our GCE setup, specifically the assumption of a constant inflow rate, which we will discuss more in the following. In the resulting model ensemble, the $k = 1.0$ category is by far the most populated, with substantially fewer cases in the other two categories. 

Figure \ref{fig:parameter_space} shows the parameter distributions of the models in three slope categories for our five free parameters: $m_{0, \mathrm{cold}}$, initial mass of the cold gas reservoir; $f_\mathrm{SNIa}$, fraction of white dwarfs with initial mass between $(3, 8)$ $\mathrm{M_\odot}$ turing into Type Ia SNe; $t_\mathrm{cool}$, cooling timescale of warm ISM; $\epsilon_\mathrm{SF,before}$, SFE during the initial prototype phase before [$\alpha$/Fe]-rise, and $\dot{m}_\mathrm{inflow,before}$. The diagonal panels show the one-dimensional distributions and the off-diagonal panels show the pairwise two-dimensional distributions. For the first column, models with $k = 1.0$ and $k = 1.4$ have similar values of $m_{0, \mathrm{cold}}$, while those with $k = 0.6$ start with a significantly more massive gas reservoir. $m_{*, cold}$ is positively correlated with $\dot{m}_\mathrm{inflow,before}$ and negatively correlated with $\epsilon_\mathrm{SF,before}$. For the second column, the distributions of $f_\mathrm{SNIa}$ are indistinguishable among the three categories. As for $t_\mathrm{cool}$, models with $k = 1.0$ require more efficient cooling than the other categories. $\epsilon_\mathrm{SF,before}$ increases with $k$ and most models have low $\dot{m}_\mathrm{inflow,before}$ of less than 5 $\mathrm{M_\odot}$ per year. 

In our GCE framework, two factors control the shape of the resulting MDF. The first is the rate of chemical evolution, [Fe/H] in this particular case, and the second is the SFH. When chemical evolution stagnates or star formation intensifies, more stellar bins are born with a certain metallicity value, flattening the slope at low metallicities or steepening the slope at high metallicities. Therefore, when models start with a large amount of gas, they enjoy a faster rate of chemical evolution and require more pristine inflow gas to either sustain a relatively high SFR or slow down chemical evolution in order to maintain the same slope. The same logic holds for SFE, which explains the correlations among the parameters. 

We translated the parameters in Fig.~\ref{fig:parameter_space} into global properties over time under our GCE framework in Figure \ref{fig:proto_mw_gas_sfh_mdf}, to illustrate the conditions associated with each MDF slope. Panels (a)--(e) show: (a) total cold+warm ISM mass; (b) star-formation rate (SFR); (c) cumulative stellar mass surviving to the present; (d) stellar $\mathrm{[Fe/H]}$; and (e) stellar $\mathrm{[Mg/Fe]}$. Consistent with Fig.~\ref{fig:parameter_space}, models with $k=0.6$ maintain an almost constant gas reservoir by starting with a large initial gas mass, receiving little subsequent inflow, and adopting a low SFE. An early star formation burst produces many stars with $\mathrm{[Fe/H]}<-2.0$, inflating the low-metallicity tail. That burst also heats cold ISM into the warm phase and quickly suppresses the SFR; the reduced late-time enrichment keeps the high-metallicity end underpopulated, yielding a shallow overall slope. 

The primary distinction between the $k=1.0$ and $k=1.4$ models is SFE. The $k=1.4$ models have substantially higher SFE, achieve the fastest early chemical evolution, and exhibit a pronounced burst even with only a moderate initial gas reservoir. The combination of high SFR and rapid enrichment steepens the metal-poor tail, producing the large $k$. A limitation of our current GCE setup is the assumption of constant inflow beginning at $t=0$ and the requirement that models reach $\mathrm{[Fe/H]}=-1.3$ by one Gyr. For the steep class ($k=1.4$), a truly constant slope would demand minimal star formation at the lowest metallicities; our target for $\mathrm{[Fe/H]}$ and early onset of gas inflow necessarily drive substantial early star formation and thus flattens the extreme metal-poor end. Conversely, a persistently shallow tail ($k=0.6$) would prefer little late-time star formation, but the constant inflow unavoidably grows the gas reservoir and steepens the high-metallicity end. These behaviors point to the need for time-dependent inflow and/or SFE in future iterations to sustain single-slope behavior over the full $-3<\mathrm{[Fe/H]}<-1.8$ range. Nevertheless, the constant inflow rate is sufficient for us to produce the desired slopes over much of our metallicity range.

Figure~\ref{fig:cum_stellar_mass_vs_feh} shows the cumulative stellar mass as a function of metallicity for the three slope classes in our GCE models (Fig.~\ref{fig:proto_mw_mdf}). We construct this diagnostic because ages for the oldest stars are notoriously uncertain, whereas ${\rm[Fe/H]}$ is directly observable at scale from Gaia XP. As a reference point, we mark the estimated stellar mass of the in-situ proto-Galaxy (the Aurora population) from \citet{2024arXiv241022250K}, who found a mass between $1.07 \times 10^7$ and $2.11 \times 10^7\,\mathrm{M_\odot}$ for stars in the range $-3.0 < \mathrm{[Fe/H]} < -2.0$. This mass was derived from modeling the spatial density distribution of metal-poor red giant branch stars from Gaia, accounting for the survey's selection function. 

The curve shapes encode the same information highlighted in Fig.~\ref{fig:proto_mw_gas_sfh_mdf}: in the burst-quenched case ($k=0.6$), an early starburst builds $\sim10^{7}\,{\rm M_\odot}$ of stars rapidly, but the large initial gas reservoir keeps the initial metallicity low, producing a fast early rise that soon plateaus. In contrast, the slow-assembly case ($k=1.4$) begins with a milder burst into a less massive reservoir, reaching a higher initial metallicity and then growing steadily under sustained low-level inflow and higher SFE; the most common slopes in both GCE models and Auriga galaxies ($k\simeq1.25$) lie between these limits. The absolute normalization of $M_\star(<{\rm[Fe/H]})$ depends on assumptions such as low-mass cutoff of the IMF. Lowering $m_{\rm min}$ from $0.5$ to $0.1\,{\rm M_\odot}$ roughly doubles the surviving stellar mass, shifting the mass-metallicity relations in Fig.~\ref{fig:cum_stellar_mass_vs_feh} upward at fixed MDF shape. However, $k=1.0$ and $k=1.4$ are still the preferred scenarios within the observed mass range. In short, the cumulative mass-metallicity plane provides an age-independent discriminator of proto-Galaxy histories and a direct bridge between XP-based MDFs, GCE models, and mass measurements from spatial modeling.

\subsection{Quiet Dawn Leads to Unified Slope}

We looked into the simulated galaxies in the Auriga project to verify the connection between the physical conditions and MDF slopes from our GCE models. The Auriga Project is a suite of high-resolution cosmological ``zoom-in'' simulations of Milky Way-mass galaxies (augmented recently to include a dwarf-galaxy sample), evolved with the moving-mesh code \textsc{Arepo} and ideal MHD to follow the coupled growth of dark matter, gas, stars, and black holes in a $\Lambda$CDM universe \citep{10.1093/mnras/stx071, 10.1093/mnras/stae1598}. 

Auriga adopts \textsc{Arepo}'s Galilean-invariant, Voronoi moving-mesh hydrodynamics--crucial for accurately capturing shocks, contact discontinuities, and rotating disks--and builds on a comprehensive galaxy-formation model (cooling, star formation, stellar/AGN feedback, metal enrichment) inherited from the Illustris framework \citep{2010MNRAS.401..791S, 2013MNRAS.436.3031V}. The simulations self-consistently amplify magnetic fields and produce disk galaxies whose global properties (e.g., H{\sc i} distributions, kinematics, scaling relations) compare well with observations, providing a physically grounded laboratory for disk formation and chemo-dynamical studies \citep{2017MNRAS.466.3859M, 2017MNRAS.469.3185P}. A recent overview details the augmented suite and the first public data release--raw snapshots, group catalogs, merger trees, and analysis tools\footnote{\url{https://wwwmpa.mpa-garching.mpg.de/auriga/data.html}}--enabling community access to the methodology and outputs \citep{10.1093/mnras/stae1598}. 

Figure~\ref{fig:auriga_galaxy_global_properties} shows the time evolution of global properties for the 30 Auriga galaxies in the original release, in the same style as Fig.~\ref{fig:proto_mw_gas_sfh_mdf}. Panels display: (a) total gas mass; (b) SFR; (c) cumulative stellar mass; (d) mass-weighted stellar [Fe/H]; and (e) mass-weighted stellar [Mg/Fe]. The commonly used Milky Way analogues-Au-17, Au-18, Au-23, and Au-26-are highlighted in orange; all other galaxies are shown in gray. Au-17 and Au-18 best reproduce the Milky Way bulge/bar chemo-kinematics and exhibit quiescent merger histories since $z\!\sim\!3.5$ \citep{2020MNRAS.494.5936F}. Au-18 also experiences an early Gaia-Sausage/Enceladus-like merger and has halo properties closest to the Milky Way \citep{2024MNRAS.531.1520M}. Au-17 also undergoes a similar merger, but its properties are less consistent with those of GSE \citep{2019MNRAS.484.4471F,2025arXiv250909576O}. 

Au-18, Au-23, and Au-26 possess Milky Way-like bars and have been used to time the disk/bar formation in our Galaxy \citep{2024MNRAS.535.2873Z}. To align the simulation with the stellar disk, the z-axis is defined by the net angular momentum vector of all stars within 0.1 times the virial radius ($< 0.1 \times r_{200c}$). We restricted the analysis to stars with $R<10$~kpc and $|\mathrm{z}|<3$~kpc to match the spatial footprint of Gaia XP catalogs. Comparing Figs.~\ref{fig:proto_mw_gas_sfh_mdf} and \ref{fig:auriga_galaxy_global_properties}, most Auriga systems are consistent with the $k\!=\!1.0$ GCE category, characterized by negligible initial gas reservoirs and minimal initial star formation. For each galaxy, we constructed MDFs from the masses and [Fe/H] of surviving star particles at $z\!\sim\!0$ and estimated the slope of the metal-poor tail over $-3 < \mathrm{[Fe/H]} < -1.8$. 

Figure~\ref{fig:Auriga_galaxy_hist_mdf_slopes} compares the slope distributions: Auriga galaxies cluster around $k \approx 1.25$, whereas the GCE models span a broader range, $k \approx 0.5$--$2.0$. The MDFs in Auriga galaxies are impacted by young metal-poor stars because many galaxies have multiple mergers bringing in metal-poor gas over time. The slopes of many Auriga galaxies would be lower when these young metal-poor stars are removed. Nevertheless, the slopes in Auriga galaxies are consistent with what we expect from the GCE models based on the gas inflow history and SFH. The original Auriga suite does not explicitly sample initial conditions with a large, pre-assembled cold-gas reservoir that ignites an early star formation burst. Because the zoom-ins draw gas from cosmological inflow and assemble disks gradually, star formation peaks happen later typically from merger- or compaction-driven. 




\section{discussion}
\label{sec:discussion}

\subsection{Selection function considerations}

{The Gaia XP metallicity catalogs considered in this work do not share a single selection function. In addition to the native Gaia XP magnitude limit, each catalog has its own effective selection through the choice of training set, inference domain, and recommended quality cuts. Throughout this work, we therefore do not impose additional cuts of our own beyond the published recommended quality cuts; our goal is to compare the public catalogs within their validated regimes rather than construct a new common sample with a different and difficult-to-model selection function.}

{A related issue is evolutionary state. Because metal-poor giants are intrinsically brighter than dwarfs, giant-dominated catalogs probe larger distances and a different mixture of Galactic components than catalogs that include substantial numbers of dwarfs and subgiants. We do not apply a uniform giant/dwarf cut across catalogs, since several public releases are already restricted to vetted giant samples whereas others are intended to operate over broader CMD regions. Instead, we assess the effect of this heterogeneity empirically. As shown above, the logarithmic MDF slope remains broadly consistent across catalogs with different CMD footprints and also across bins of observed absolute magnitude and Galactocentric radius (Figs.~\ref{fig:gaia_xp_mdf}, ~\ref{fig:gaia_xp_mdf_Gobs}, and~\ref{fig:gaia_xp_mdf_R}) indicating that luminosity-class and volume effects primarily alter the normalization and sampled volume rather than strongly changing the fitted MDF slope over the adopted metallicity interval.}

{Disk contamination is another concern because the Gaia XP samples span both low- and high-$|\mathrm{z}|$ regions. We do not apply a global kinematic cut to the catalogs because only a minority of stars ($\sim$5--10\%) have Gaia DR3 RVS radial velocities; instead, we use this subset only as a diagnostic. The available kinematics show that stars with $\mathrm{[Fe/H]}<-2.0$ are predominantly slow-rotating, consistent with a proto-Galactic population, whereas the clearest fast-rotating contaminants become common closer to the metal-richer boundary, especially above $\mathrm{[Fe/H]}\gtrsim-1.5$ (Fig.~\ref{fig:gaia_xp_feh_vphi}). These stars often deviate from the old, metal-poor isochrone locus in the CMD and occupy regions that are sparsely represented in the training sets of several XP-based pipelines (Fig.~\ref{fig:gaia_xp_hr_vphi}), indicating that they are more likely metal-rich disk contaminants with underestimated metallicities than genuine very metal-poor stars with large angular momentum. Catalogs with larger low-$|z|$ samples are therefore more susceptible to this effect (see the last panel of Fig.~\ref{fig:gaia_xp_feh_vphi} and Fig.~\ref{fig:gaia_xp_R_z}). To minimize such contamination, we restrict the MDF-tail analysis to $\mathrm{[Fe/H]}<-2.0$ and treat the RVS-based kinematic distributions only as a validation check rather than as a selection criterion. The confinement of the clearest contaminants to the metal-richer boundary suggests that their impact on the fitted MDF slope over $-3<\mathrm{[Fe/H]}<-2.0$ is small.}

\subsection{A Unified Proto-Galactic Population}

{We interpret the proto-Galaxy as the composite stellar population that formed up to the transition from dispersion-dominated assembly to the emergence of a rotation-supported disk. Across the Gaia XP metallicity catalogs, the metal-poor MDF tails over $-3<\mathrm{[Fe/H]}<-2.0$ are generally well described by an approximately single-slope exponential with logarithmic slopes centered near $k\approx1$. After applying each catalog's recommended quality cuts, we find no compelling evidence for secondary peaks or shoulders within this interval. Residual departures appear mainly near the boundaries and are consistent with known pipeline and selection effects: APOGEE-anchored products tend to steepen below $\mathrm{[Fe/H]}\approx-2$ because of sparse metal-poor training labels, whereas VMP-optimized or CaHK-based designs can flatten the extreme metal-poor end. These effects likely account for much of the inter-catalog variation in the fitted slopes, but they do not qualitatively change the main result that the metal-poor tail is broadly smooth and close to exponential over the adopted metallicity range.}

Taken together, these pipeline and selection effects explain the apparent differences among catalogs without requiring genuine departures from a single-exponential halo MDF over $-3\le\mathrm{[Fe/H]}<-2.0$. This result holds even when MDFs are reexamined in Galactocentric-radius or magnitude slices. Residual differences track known selection footprints-XP spectra exist only for $G<17.65$, and some catalogs restrict inference to specific luminosity classes (e.g., vetted RGB), altering CMD occupancy and the inner/outer-halo mixture. Within these limits, the slope stability across $R$ and magnitude supports a globally similar metal-poor tail.

Both observations and simulations show that such old, metal-poor stars arise from a mixture of origins: an in-situ component heated by early mergers and accreted debris from one or two major progenitors \citep{2010ApJ...708.1398T,2018MNRAS.480..652E,2024MNRAS.527.9810H}. However, disentangling these channels is difficult even in simulations with perfect birth tags-kinematics and chemistry overlap strongly below $\mathrm{[Fe/H]}<-1$, with mean abundance offsets $\lesssim0.3$ dex and broad MDF overlap in both prograde and retrograde regions \citep{2021MNRAS.503.5846R,2023A&A...677A..91K}. Observational separations based on integrals of motion recover only a small subset of accreted structures \citep{2022ApJ...937...14B}, and recent FIRE-2 analyses show significant kinematic and chemical blending during the proto-Galactic phase \citep{2024MNRAS.527.9810H}.

Given this intrinsic overlap, we adopt a pragmatic approach: rather than attempting to classify individual stars, we treat the proto-Galaxy as a single, old ($\gtrsim11$ Gyr), metal-poor composite population and infer its aggregate physical state-its early gas reservoir, star-formation efficiency, and inflow history. If the early Milky Way was dominated by one main progenitor, these quantities represent its intrinsic properties; if multiple comparable systems contributed, our constraints should be interpreted as their mass-weighted averages. In either case, the chemical continuity of the metal-poor MDF supports a unified treatment of all stars old enough to record the conditions of the proto-Galaxy.

\subsection{Addressing Caveats on MDFs}

A systematic uncertainty arises from attenuation bias--the tendency of different inference methods to distort the intrinsic shape of the metal-poor MDF. Template-fitting and synthetic-spectra approaches that explicitly include very metal-poor stars (e.g., Ca\,H\&K-based or VMP-augmented training) extend sensitivity into the lowest metallicities, but also introduce broader uncertainties that flatten the observed tail, yielding shallower slopes ($k_{\mathrm{obs}} < k_{\mathrm{true}}$). In contrast, catalogs derived by label transfer from APOGEE or other spectroscopic surveys suffer from a lack of true VMP training examples, even when augmented with a few LAMOST or Pristine stars, causing the model to compress the dynamic range and overpredict metallicities at the low end. 

This truncation steepens the inferred tail ($k_{\mathrm{obs}} > k_{\mathrm{true}}$), mimicking the effect of an intrinsically smaller gas reservoir or higher star-formation efficiency in GCE interpretations. The net impact of attenuation therefore depends on the inference framework, the effective [Fe/H] prior, and the treatment of noise and regularization. Cross-catalog comparisons must thus account for this methodological bias before attributing differences in $k$ to physical variations in the proto-Galaxy. A more detailed and careful intra-catalog examination is required in the future to further narrow down the slope of the MDF tail. 

A related consideration concerns the implicit selection function of stars from the GCE models and its connection to the observed CMD coverage of Gaia XP catalogs. The MDF represents stars that have survived to the present day, effectively sampling the least massive, longest-lived stars. In principle, differences among catalogs in their CMD coverage--for example, whether they focus on main-sequence dwarfs or red-giant branches--could bias the inferred MDF if stars of different masses contributed differently to the metal-poor tail. However, inspection of the GCE model outputs shows that nearly all stars surviving to $z=0$ originate from one or a few of the lowest-mass bins in the adopted IMF ($\mathrm{M}_* < 0.7 \mathrm{M}_\odot$), since higher-mass stars have long evolved and recycled their metals into the ISM. 

Consequently, the shape of the MDF is governed almost entirely by the cumulative star formation and enrichment history rather than by the present-day stellar mass distribution. Because all XP-based catalogs ultimately probe these same long-lived low-mass populations, their differing CMD footprints do not systematically affect the inferred slope $k$. In this regime, the MDF can be interpreted as a faithful tracer of the proto-Galaxy's early gas-regulation history, largely insensitive to catalog-specific evolutionary or photometric selections. {Nevertheless, we implemented different selection functions based on Gaia magnitudes of the stellar component in our models and found that the resulting MDFs are identical. This agrees with Fig.~\ref{fig:gaia_xp_mdf_Gobs} in which MDF slopes remain consistent across different parallax-corrected magnitude bins for catalogs with sufficient stars.}    

\subsection{Comparison with Previous Halo MDF Studies}

The slopes inferred from Gaia XP metallicity catalogs are broadly consistent with previous constraints on the Milky Way's metal-poor tail. In the Pristine survey, the inner-halo MDF between $-3.4<\mathrm{[Fe/H]}<-2.5$ follows a single exponential with $k\equiv\Delta\log_{10}N/\Delta\mathrm{[Fe/H]}=1.0\pm0.1$ for main-sequence turnoff stars, establishing a canonical baseline of $k\simeq1$ \citep{2020MNRAS.492.4986Y}. SkyMapper photometric metallicities for $\sim10^5$ giants yield two regimes: $k=1.5$ for $-3.0<\mathrm{[Fe/H]}<-2.3$ and $k\simeq1.0$ for $-2.3<\mathrm{[Fe/H]}<-1.5$, with negligible dependence on Galactic height to $|\mathrm{z}|\approx5$--7~kpc \citep{2019MNRAS.489.5900D,2021ApJ...911L..23C}. These values bracket the range found here from Gaia XP, confirming that the inner-halo MDF tail typically follows an exponential decline with $k\approx1$.

Other halo MDF studies, with differing spatial coverages and selection functions, have emphasized the metal-poor tail in the outer halo. The Hamburg/ESO Survey revealed a sharp decline near $\mathrm{[Fe/H]}\approx-3.6$, interpreted as evidence for a critical metallicity $Z_\mathrm{cr}\approx10^{-3.4}Z_\odot$ required for low-mass star formation \citep{2007MNRAS.381..647S,2009A&A...507..817S}. SDSS Stripe-82 and BOSS spectroscopy described a multi-modal halo MDF with peaks near $-1.7$ and $-2.3$, corresponding to inner- and outer-halo components with distinct kinematics \citep{2013ApJ...763...65A,2014A&A...568A...7A}. LAMOST giants at $|\mathrm{z}|>5$~kpc show a similar three-component structure-inner and outer halo plus a heated-disk population \citep{2018ApJ...862..163L}. The H3 Survey combined with Gaia data indicates that $\gtrsim95\%$ of halo giants can be attributed to discrete merger remnants: beyond 15~kpc, the halo is dominated by GSE and Sagittarius debris, while the most metal-poor tail arises from lower-mass progenitors \citep{2020ApJ...901...48N,2021A&A...651A..79B}.

In contrast to these high-$|\mathrm{z}|$ or accretion-dominated samples, nearly all Gaia XP stars lie within $|\mathrm{z}|<2$~kpc (Fig.~\ref{fig:gaia_xp_R_z}), probing the chemically mixed, in-situ inner halo. The exponential slopes derived here therefore trace the early enrichment history of the proto-Galactic gas reservoir rather than the outer-halo debris field. Methodologically, the all-sky XP spectroscopy provides a more uniform metallicity baseline and higher completeness at low $|\mathrm{z}|$ than prior spectroscopic surveys, reducing spatial biases in the inferred MDF shape. Physically, the near-unity slope $k\approx1$ is consistent with a moderately efficient early star-formation episode within a well-mixed gas reservoir-close to the expectation of a one-zone, leaky-box model with limited metal retention. Steeper slopes ($k>1$) observed in some photometric or high-$|\mathrm{z}|$ samples likely reflect either stronger outflows or incomplete sampling of the metal-poor tail.

{\citet{2025PASA...42...40L} provided an independent check on the metal-poor MDF tail using follow-up spectroscopy of Gaia XP–selected candidates. They selected $\sim$6000 Gaia DR3 candidates from the \citet{2023MNRAS.524.1855Z} catalog, obtained 2dF+AAOmega spectroscopy, and then measured MDF tails for kinematically defined components (halo, prograde/retrograde disk, and GSE). In their Fig.~16, the fitted tail slopes over $-3\le\mathrm{[Fe/H]}\le-2$ are close to unity (for the halo they report $\Delta\log N/\Delta\mathrm{[Fe/H]}=1.15\pm0.08$), consistent with the $k\approx1$ slopes we found from Gaia XP MDF tails over the same metallicity range. In their appendix~A, they compared their spectroscopic metallicities to several Gaia XP-based catalogs and found that the level of agreement is strongly pipeline-dependent, with substantial scatter and clear systematics at the metal-poor end. In particular, they show that XP products by \citet{2023MNRAS.524.1855Z},  \citet{2023ApJS..267....8A}, and \citet{2024ApJS..272....2L} tend to assign systematically higher $\mathrm{[Fe/H]}$ to their most metal-poor stars, plausibly driven by reddening-related parameter degeneracies and limited metal-poor coverage in the training sets. This systematic over-estimation of $\mathrm{[Fe/H]}$ at the metal-poor end would shift intrinsically VMP/EMP stars toward higher inferred metallicities, thereby steepening the observed decline. The true metal-poor MDF tail would therefore be flatter (i.e., have a smaller $k$) than that measured from the affected catalogs.}

\subsection{Comparison with Previous GCE Studies}

{The observed rise in [$\alpha$/Fe] provides a key boundary condition for our GCE models introduced in Paper~I, ruling out many implausible parameter combinations. The observed MDFs presented here further constrain the models by breaking degeneracies among parameters. As shown in Fig.~\ref{fig:Auriga_galaxy_hist_mdf_slopes}, the median slope from our original model grid already agrees well with the median slope of the Auriga galaxies, and the overall slope distribution in the left panel matches the observed MDF slopes from the Gaia XP metallicity catalogs in Fig.~\ref{fig:gaia_xp_mdf}. In Paper~I, we found that the [$\alpha$/Fe] rise alone sets only partial constraints: it defines a lower limit on the SFE and upper limits on the inflow rate and warm-ISM cooling timescale but does not meaningfully restrict the initial gas reservoir mass (see Fig.~7 in Paper~I). It primarily constrains the SNe~Ia fraction to $\sim6\%$. The inclusion of the MDF tail from Gaia XP now allows us to tighten these limits substantially. For the most common slope category ($k = 1.0$), the feasible parameter ranges correspond to an initial cold-gas reservoir below $10^8\,\mathrm{M_\odot}$, a star-formation efficiency of a few~$\times10^{-10}\,\mathrm{yr}^{-1}$, and an inflow rate of approximately $2\,\mathrm{M_\odot}\,\mathrm{yr}^{-1}$.}

Classical one-zone models show that the low-metallicity tail of the halo MDF should be exponential, with the power set by either lowering the effective yield (mass of freshly produced metals returned to the ISM per unit mass locked in long-lived stars/remnants) by a factor of 13 compared to the disk or adding outflow proportional to star formation \citep{1976ApJ...209..418H}. This simple model successfully predicted the number of metal-poor stars down to $\mathrm{[Fe/H] = -3}$ for early observations \citep{1991AJ....101.1865R, 1996AJ....112..668C}. However, \citet{2003A&A...404..211P} noted that this simple model over-predicts the number of stars below $\mathrm{[Fe/H] < -4}$ if instantaneous recycling approximation (IRA) is relaxed and found that a non-IRA model with a moderate initial inflow and star-driven outflow recovers the observed exponential tail without significant reduction in the effective yield. 

In a hierarchical extension, the halo MDF can be written as the sum of subhalo MDFs with mass-dependent yields \citep{2008A&A...489..525P}. In this picture, the low-metallicity bins ($\mathrm{[Fe/H] < -2.0}$) are supplied primarily by numerous low-mass stellar subhaloes ($10^{6-7}~\mathrm{M_\odot}$). Independently, a merger-tree GCE model finds that the halo MDF requires strong feedback plus pre-enriched infall \citep{2007MNRAS.381..647S}. In their fiducial solution, 60\% of the metals that pre-enrich the Galactic medium are ejected by halos with a dark matter mass less than $6 \times 10^9~\mathrm{M_\odot}$. Finally, a two-infall GCE model augmented with star-driven outflow reproduces the halo MDF for with a mass-loading factor of 14 and an early infall timescale less than 0.2 Gyr, implying rapid early enrichment and efficient metal loss \citep{2013A&A...554A.135B}. These studies depict an open, rapidly evolving proto-Milky Way regulated by gas flows and hierarchical assembly. An early infall episode coupled to outflows enables rapid enrichment while losing a large amount of metals. The lowest-metallicity stars chiefly arise in numerous low-mass progenitors whose ejecta pre-enriched incoming gas. 

We employ a grid of one-zone, non-IRA GCE models that vary the initial gas reservoir, SNe Ia fraction, warm-ISM cooling timescale, SFE, and gas inflow rate over physically motivated ranges. We group outcomes by the exponential metal-poor-tail slope $k$. The models include outflows proportional to SFR and supernovae via mass-loading factors and were pre-selected in Paper~I to achieve a characteristic turn in $[\mathrm{Fe/H}]$ -$[\mathrm{Mg/Fe}]$ within the first 1--2 Gyr, consistent with the observed proto-Galactic timeline. These models typically favors small gas reservoirs during the first Gyr. The models are capable of reproducing a single-slope exponential MDF with $k \simeq 1$, while departures reflect the simplifying assumption of a constant inflow rate. 

In our framework, $k$ traces the early gas regulation: shallow tails ($k\simeq0.6$) arise in models with large initial reservoirs and little subsequent inflow, whereas steeper tails ($k\simeq1.0$--$1.4$) arise for smaller initial reservoirs plus sustained inflow that builds the gas reservoirs while delaying enrichment. The trade-off between SFE and gas cooling also modulate the mapping from the parameter space to MDF unless one parameter takes an extreme value. Fitting the same exponential form to local Auriga star-particle MDFs ($R < 10~\mathrm{kpc}$, $|\mathrm{z}| < 3~\mathrm{kpc}$) yields a distribution clustered near $k\simeq1.25$, consistent with a gas-regulated, infall-driven growth scenario under our framework, which we confirm through the global properties of Auriga galaxies. This scenario is also supported by more detailed examination of the gas accretion history of Auriga galaxies by \citet{2022MNRAS.517..832I}. This picture is compatible with classical one-zone interpretations in which the tail slope traces the effective yield. A moderate initial gas reservoir of $10^{8-9} \mathrm{M_\odot}$ aligns with early-infall models \citep{2003A&A...404..211P,2013A&A...554A.135B}; cooling timescales need only be moderately short-no extreme suppression of the effective yield is required-and low-to-moderate inflow rates are consistent with pre-enrichment by numerous low-mass progenitors \citep{2007MNRAS.381..647S,2008A&A...489..525P}.

\subsection{The Case for a Burstless Proto-Galaxy}

The MDF constraints from Gaia XP present a case for a largely burstless proto-Galaxy. While cosmological zoom-in simulations frequently show that Milky Way–mass progenitors at $z\gtrsim2$ undergo bursty star formation, our GCE analysis shows that the observed, near-unity slope ($k \approx 1.0$) is inconsistent with a significant early starburst. A shallow MDF tail ($k \simeq 0.6$), the expected signature of a massive initial burst, is not observed. Instead, the exponential decline is best reproduced by models with sustained, low-level inflow and star formation.

Cosmological zoom-in simulations show that Milky Way-mass progenitors at $z\gtrsim2$ undergo bursty star formation. In the cold-mode regime of \citet{2006MNRAS.368....2D}, efficient cooling prevents a stable virial shock for halos with $M_{\rm vir}\lesssim10^{12}\,{\rm M_\odot}$ at $z\gtrsim2$, so dense, cold inflow drives recurrent bursts. In FIRE, these bursts are followed by energetic winds with mass loading $\eta\sim10$ that expel a substantial fraction of the ISM and transiently depress the SFR \citep{2015MNRAS.454.2691M}. In VINTERGATAN, high-$z$ bursts arise from a turbulent, cold ISM reacting to frequent mergers \citep{2022MNRAS.516.2272S}. In FIRE-2, the bursty phase ends as a quasi-static, hot inner CGM develops when the halo crosses $M_{\rm vir}\sim10^{12}\,{\rm M_\odot}$ \citep{2021ApJ...911...88S}, accompanied by a rapid ($\lesssim1\,\mathrm{Gyr}$) transition to time-steady, rotation-supported disk star formation \citep{2023MNRAS.519.2598G}. Throughout the proto-galaxy phase, the ISM remains turbulent and regulated by coupled inflow-outflow cycles; feedback-driven outflows throttle the net supply to the nascent disk until the accretion mode changes, but do not fully quench star formation \citep{2025ApJ...990....7S}.

This early, burst-dominated mode plausibly overlaps the early epoch of disk formation \citep{2021MNRAS.505..889Y}. Kinematic diagnostics indicate that dispersion-dominated (low-circularity) orbits are characteristic of stars born during the proto-galaxy and early disk phase \citep{2023MNRAS.523.6220Y,2024MNRAS.527.6926M}, and stars formed in these early bursts can contribute to observed properties at $z \sim 0$ attributed to merger debris such as GSE \citep{2023MNRAS.521..995R}. In our GCE framework, a shallow metal-poor tail ($k\approx0.6$) arises from a large initial star-formation burst followed by ISM heating and weak subsequent inflow, whereas sustained gas accretion and weaker initial bursts generally yields steeper tails ($k\gtrsim1$). MDFs inferred from Gaia XP catalogs appear smooth-lacking obvious multi-modal structure like the episodic features modeled by \citet{2025OJAp....8E...7T}-but measurement uncertainties can wash out burst-scale signatures, especially since the surviving stellar mass in an individual burst is less than $ 10^7 \ \mathrm{M_\odot}$. \citet{2024arXiv241022250K} estimated the stellar mass of the in situ ``Aurora'' and GSE to be $\sim 3 \times 10^7 \ \mathrm{M_\odot}$ between $-3 < \mathrm{[Fe/H] < -2.0}$ which would prefer a steeper slope ($k \simeq$ 1.0--1.4) based on Fig.~\ref{fig:proto_mw_gas_sfh_mdf} c) and ~\ref{fig:cum_stellar_mass_vs_feh}.

\section{Conclusion}
\label{sec:conclusion}
The chemical and kinematic structure of the oldest stars encodes the conditions under which the Milky Way first assembled. In our Paper I, we showed that the moderate rise in [$\alpha$/Fe] near $\mathrm{[Fe/H]} \approx -1.3$ is best explained by the onset of $\alpha$-enhanced inflow that supplied both fuel and angular momentum as the Galactic disk began to form. This feature identified the transition from the turbulent, dispersion-dominated Aurora phase to a rotation-supported system, but its amplitude alone could not uniquely constrain the proto-Galaxy's gas reservoir, inflow rate, or star-formation efficiency. 

In this paper, we extend that analysis by turning to an independent diagnostic, the slope of the metal-poor tail of the Milky Way's MDF, which encodes the cumulative balance between metal production and gas supply during the earliest phase of Galactic evolution. Using metallicities from multiple Gaia DR3 XP-based catalogs, we quantify the logarithmic slopes of the MDF over $-3<\mathrm{[Fe/H]}<-2.0$ and interpret them with a grid of one-zone, non-IRA GCE models calibrated to reproduce the [$\alpha$/Fe] rise. This approach links the metal-poor MDF directly to the physical regulation of the proto-Galactic gas reservoir and provides a complementary constraint to abundance-ratio-based inferences. 

Our main results are:
\begin{enumerate}
    \item Kinematic diagnostics confirm that stars with $\mathrm{[Fe/H]}\lesssim-1.5$ exhibit negligible net rotation, with median $v_\phi\approx0~\mathrm{km\,s^{-1}}$, consistent with expectations for the dispersion-dominated proto-Galactic population. The low scatter in $v_\phi$ across Gaia XP catalogs indicates that these stars belong to an old, dynamically hot component rather than a mixture of rotating substructures. An apparent excess of stars with $v_\phi>200~\mathrm{km\,s^{-1}}$ near $\mathrm{[Fe/H]}\gtrsim-1.5$ (Fig.~\ref{fig:gaia_xp_feh_vphi}) is not a true halo feature but arises from thin-disk contaminants whose metallicities are underestimated in certain inference pipelines, particularly those trained primarily on red-giant stars or limited CMD domains.

    \item {MDFs from all Gaia XP catalogs, after the recommended quality cuts, are broadly consistent with single-slope exponentials over $-3<\mathrm{[Fe/H]}<-2.0$, with slopes centered near $k\approx1$ and an interquartile range of 0.6--2.0 (Figs.~\ref{fig:gaia_xp_mdf}, \ref{fig:gaia_xp_mdf_R}, and \ref{fig:gaia_xp_mdf_Gobs}). The slope remains largely invariant with Galactocentric radius ($R\lesssim10$ kpc) and with parallax-corrected magnitude, indicating limited qualitative sensitivity to spatial incompleteness or luminosity-class selection. Modest inter-catalog differences follow known design choices: APOGEE-anchored label-transfer pipelines yield slightly steeper tails ($k\gtrsim1.2$) owing to limited VMP training coverage, whereas VMP-optimized or CaHK-augmented models produce flatter tails ($k\lesssim0.8$). These catalog-dependent effects plausibly explain much of the observed spread in $k$, but they do not alter the main conclusion that the Gaia XP metallicity catalogs consistently recover a broadly smooth, approximately exponential metal-poor tail.}

    \item Our grid of one-zone GCE models quantifies the direct link between the MDF tail slope and the proto-Galaxy's gas-regulation parameters.
    Shallow tails ($k\simeq0.6$) occur only when the initial cold gas reservoir exceeds $\sim10^9\,\mathrm{M_\odot}$, the SFE is below $10^{-10}\,\mathrm{yr^{-1}}$, and inflow contributes less than $20\%$ of the initial mass over the first Gyr. 
    Such models produce an early starburst that rapidly heats and displaces the ISM, leading to a steep decline in the SFR.
    Conversely, steeper tails ($k\simeq$1.2--1.4) arise for smaller initial reservoirs ($\lesssim10^8\,\mathrm{M_\odot}$) with sustained, low-level inflow and SFE of order $10^{-9}\,\mathrm{yr^{-1}}$, indicative of gradual, infall-driven growth (Figs.~\ref{fig:parameter_space} and \ref{fig:proto_mw_gas_sfh_mdf}).
    The Gaia XP-based slopes cluster near $k\simeq1.0\pm0.2$, implying a proto-Galaxy with a moderate initial gas mass of $10^{8-9}\,\mathrm{M_\odot}$, moderate inflow, and non-extreme SFE--consistent with a ``quiet'' early buildup rather than a strongly burst-dominated or inflow-starved phase.
    Complementary diagnostics such as the cumulative stellar mass-metallicity relation (Fig.~\ref{fig:cum_stellar_mass_vs_feh}) will further refine these constraints.

    \item Among the GCE models that reproduce the observed [$\alpha$/Fe]-rise associated with the Galactic spin-up, the modal metal-poor-tail slope is $k\simeq1.2$. In the Auriga simulation suite, Milky Way-mass analogs exhibit nearly identical slopes ($k\simeq1.25$) with little galaxy-to-galaxy scatter (Fig.~\ref{fig:Auriga_galaxy_hist_mdf_slopes}). This further demonstrates the constraining power of the observed [$\alpha$/Fe]-rise on the proto-Galaxy. These simulated systems assemble their baryons gradually over the first $\sim$1--2 Gyr without a large, isolated initial starburst, closely resembling our $k\simeq1.0$ GCE models in which gas reservoirs build slowly under low-to-moderate inflow (Fig.~\ref{fig:proto_mw_gas_sfh_mdf} and ~\ref{fig:auriga_galaxy_global_properties}). The concordance between the observed, modeled, and simulated slopes indicates a shared early phase in which a turbulent, feedback-regulated ISM throttles net accretion, naturally producing near-exponential metal-poor tails ($k\approx1 - 1.2$) prior to the onset of $\alpha$-enhanced refueling at the start of disk formation.

\end{enumerate}

Our results, derived from the chemical imprint of the gas-regulation history, provide a diagnostic of the proto-Galaxy's initial conditions that is independent of---and more robust than---kinematic signatures. This distinction is crucial because the kinematic record of the earliest phases can be overwritten by later disruptive events. \citet{2025arXiv250909576O} demonstrates that massive radial mergers like the GSE can fundamentally scramble orbital information, making the kinematics of the proto-Galaxy stars and stars kicked up from the disc (the Splash) indistinguishable at low metallicities. Their work shows that stellar kinematics at $z=0$ rarely recover the true timing of the disc's initial spin-up due to this merger-induced scrambling. In contrast, the MDF is a product of the galaxy's internal gas supply and enrichment history, which is more resilient to such dynamical shuffling. Therefore, our method based on the MDF slope offers a more direct window into the conditions of the pre-merger, proto-Galactic gas reservoir, providing complementary constraints to purely kinematic approaches.

The Gaia XP metallicity catalogs differ in their training sets, flux calibrations, and CMD masks, and residual systematics-particularly near the Galactic plane-may still affect completeness and flatten the counts at $\mathrm{[Fe/H]}\lesssim-2.5$. Our present one-zone framework, while flexible, assumes constant inflow within two one-Gyr phases and does not yet include explicitly time-variable, merger-driven accretion. Future work should incorporate a hierarchical forward model of the MDF that embeds each catalog's selection function within a multi-zone GCE framework, allowing inflow, star-formation efficiency, and feedback to vary continuously in time while marginalizing over Type Ia delay-time distributions and uncertain yields. The next generation of datasets-Gaia DR4/DR5, WEAVE \citep{2012SPIE.8446E..0PD, 2024MNRAS.530.2688J}, and 4MOST \citep{2019Msngr.175....3D}-will provide higher-fidelity kinematics and XP-based [$\alpha$/M] and [C/Fe] ratios, enabling cleaner separation of disk contaminants and improved recovery of the faintest metal-poor populations. Together, these advances will allow the exponential VMP tail to serve as a precise, physically interpretable observable linking the Milky Way's earliest gas supply, enrichment, and star-formation history to the emergence of its first long-lived disk.

\bibliography{main}{}
\bibliographystyle{aasjournal}

\end{document}